\documentclass[conference]{IEEEtran}
\IEEEoverridecommandlockouts
\usepackage{cite}
\usepackage [english]{babel}
\usepackage [autostyle, english = american]{csquotes}
\MakeOuterQuote{"}
\usepackage{amsmath,amssymb,amsfonts}
\usepackage{pifont}
\newcommand{\cmark}{\ding{51}}%
\newcommand{\xmark}{\ding{55}}%
\usepackage{amsthm}
\newtheorem{definition}{Definition}
\usepackage{algorithmic}
\usepackage{graphicx}
\usepackage{textcomp}
\usepackage{dblfloatfix}
\usepackage{url}
\usepackage{booktabs}
\usepackage{multirow}
\usepackage{comment}
\usepackage{makecell}
\usepackage{xcolor}
\usepackage{tablefootnote}
\usepackage{float}
\usepackage{pgfplots}
\usepackage{tabularray}
\usepackage{balance}

\usepackage[usestackEOL]{stackengine}
\pgfplotsset{width=7cm,compat=1.9}
\def\BibTeX{{\rm B\kern-.05em{\sc i\kern-.025em b}\kern-.08em
    T\kern-.1667em\lower.7ex\hbox{E}\kern-.125emX}}
\begin{document}

\title{Elevating Software Trust: Unveiling and Quantifying the Risk Landscape\\
}

\author{Sarah Ali Siddiqui\thanks{E-mail:sarah.siddiqui@data61.csiro.au}, Chandra Thapa, Rayne Holland, Wei Shao, and Seyit Camtepe\\
\IEEEauthorblockA{CSIRO Data61, Sydney, Australia 
}}

\maketitle


\begin{abstract}
Considering the ever-evolving threat landscape and rapid changes in software development, we propose a risk assessment framework called SAFER (Software Analysis Framework for Evaluating Risk). This framework is based on the necessity of a dynamic, data-driven, and adaptable process to quantify security risk in the software supply chain. Usually, when formulating such frameworks, static pre-defined weights are assigned to reflect the impact of each contributing parameter while aggregating these individual parameters to compute resulting security risk scores. 
This leads to inflexibility, a lack of adaptability, and reduced accuracy, making them unsuitable for the changing nature of the digital world.
We adopt a novel perspective by examining security risk through the lens of trust and incorporating the human aspect. Moreover, we quantify security risk associated with individual software by assessing and formulating risk elements quantitatively and exploring dynamic data-driven weight assignment. This enhances the sensitivity of the framework to cater to the evolving security risk factors associated with software development and the different actors involved in the entire process. The devised framework is tested through a dataset containing 9000 samples, comprehensive scenarios, assessments, and expert opinions. Furthermore, a comparison between scores computed by the OpenSSF scorecard, OWASP risk calculator, and the proposed SAFER framework has also been presented. The results suggest that SAFER mitigates subjectivity and yields dynamic data-driven weights as well as security risk scores. 
\end{abstract}


\section{Introduction}
\label{sec:intro}
In today's interconnected and digital world, the creation and distribution of software have evolved into intricate processes known as software supply chains. Similar to traditional supply chains that involve the production, distribution, and delivery of physical goods, software supply chains encompass various stages through which a piece of software is developed, tested, packaged, and ultimately delivered to consumers \cite{new12023, new22022}. These stages involve collaboration among different stakeholders, including developers, vendors, suppliers, and distributors, all working together to bring software to life \cite{new62023}
. Ensuring safe, reliable, and efficacious ways to deliver software products to end users is fundamental to the software supply chain \cite{new32011}. 

The diversity of entities engaged in the process, along with the 3rd party components (\emph{e.g.}, equipment, code, and libraries), magnifies the susceptibilities and security risks \cite{new42020, new52023}. 
According to the European Union Agency for Cybersecurity's (ENISA) report, attacks related to the software supply chain are considered the prime threat in the current threat landscape \cite{enisa2021}. 
Vulnerability in or attack on even a single component of these chains has catastrophic impacts as it has the potential to propagate to other components within the same supply chain \cite{cisa2022}. Therefore, it is of paramount importance that such security risks be identified and addressed suitably. Moreover, if there exist multiple vulnerabilities or security risks, an organization has to determine which vulnerability to address first \cite{newrisk2015}. Please note that the term \textit{risk} refers to \textit{security risk} throughout the paper.

%
When choosing software to download, users must compare multiple (risky) options and make decisions based on objective criteria rather than subjective opinion \cite{newprivrisk2020}.
%
Risk assessment guides decisions in scenarios like prioritizing resource allocation for addressing security issues or choosing safer software options~\cite{newrisk2015}.
Moreover, it is important to quantify the risk to estimate the severity of the risk and eliminate subjectivity and inconsistencies in evaluating these risk levels to make informed decisions \cite{newcarbonrisk2023}. For instance, a subjective risk assessment performed to evaluate a piece of software might yield a low-risk level when carried out by an individual, and the same piece of software might be assessed as a high-risk level by another individual in the same setting in the same organization. This subjective and inconsistent assessment makes it difficult for the organization to make decisions regarding the software.

\begin{figure*}[!b]
\centering
\includegraphics[scale=0.70]
{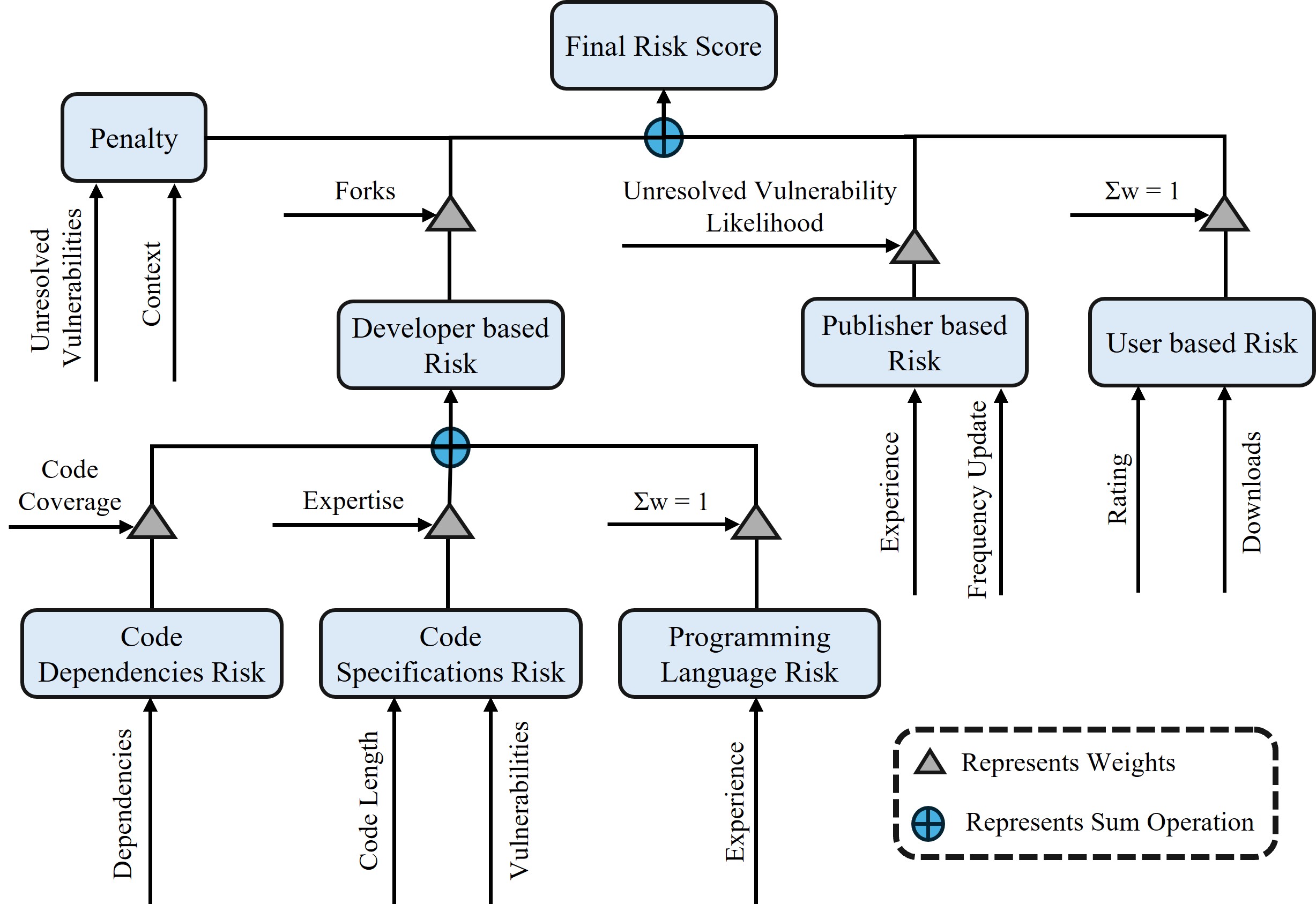}
\caption{Proposed SAFER Framework: Individual parameters are computing the actor-based risks (\emph{i.e.}, developer, publisher, and the user) and their respective weights, which are fed along with the penalty factor to compute the final risk score.}
\label{fig:tree_diag_sriqt}
\end{figure*}

Most research works on software risk assessment centers around qualitative analyses \cite{nistcvss, new92011, new10sei, new11army, new12airforce, new1392}, which introduces a factor of subjectivity and inconsistency in outcomes. For instance, if two people evaluating the risk associated with the same software have different levels of experience/expertise, the risk assessed by them will be different due to their subjective opinions. In 
terms of quantitative risk evaluation, weights are associated with individual parameters to reflect their influence on the cumulative risk score. These weights are often static, pre-defined, and there exists a lack of justification for the selection of preset values for these weights (\emph{e.g.}, if the number of parameters influencing the cumulative risk score is $p$, taking a mean of all these parametric values suggest a static and pre-defined weight of $\frac{1}{p}$ associated with each parameter without any reasoning to justify why the equal weight $\frac{1}{p}$ is employed instead of $\frac{0.5}{p}$ for one and $\frac{1.5}{p}$ for the other and so on), which also introduces an element of subjectivity. The objective quantification of these weights has been overlooked while formulating risk evaluation models \cite{nistcvss, new142011}. Moreover, a significant proportion of these risk assessment frameworks emphasize only the software development stage and accordingly employ parameters regarding only the software code itself \cite{chess, donald}. In addition, the omission of software dependencies from risk management models is a significant limitation that potentially undermines the effectiveness of these models \cite{slsa}. Furthermore, the human aspect has been ignored while defining the threat landscape for risk assessment.

The perceived reliability and security of software considerably impact the willingness of an end user to use it, i.e., if the software is susceptible to security risks, the user’s confidence in that software weakens. 
Assessing the trustworthiness of an entity helps in understanding the potential risks associated with that entity \cite{newrisktrust2021}. In other words, trust is seen as a risk mitigation strategy, and it reduces perceived risk \cite{new72022}. The existing literature indicates a noteworthy gap in examining software risk, specifically in relation to the perspective of trust. To the best of our knowledge, none of the current studies have explored software risk from the trust standpoint, highlighting a substantial and unexplored area within the research landscape. It is of significant importance to study and define trust to assess software ecosystems \cite{new152023}. 

Accordingly, we propose a novel software risk assessment framework called SAFER (Software Analysis Framework for Evaluating Risk) that relies on quantitative influencing parameters to formulate individual risk segments along with data-driven and dynamic weights associated with these risk segments to address subjectivity and inconsistency in risk assessment. It is essential to emphasize that this paper considers risk from a security point of view. 
 While referring to dynamic and data-driven weights, it is worth mentioning that the weights themselves do not impact secure software development practices but reflect/represent the characteristics of the software development and vulnerabilities. Employing data-driven weights adds to the transparency of the risk assessment by reflecting the actual threat landscape. Moreover, if an end user (with limited or no knowledge regarding software development or exposure) wishes to customize the framework, they will require a deep understanding of each and every aspect to make an informed decision regarding the selection of appropriate pre-defined static weights. Whereas, in the proposed framework, these weights are kept dynamic so they will automatically adapt to the data provided without needing a thorough understanding. SAFER also considers the human aspect of the software supply chain and incorporates parameters related to the developer, publisher, and the user in addition to software code and dependencies-related attributes from a trust perspective, \emph{i.e.}, it assesses the risk associated with the software through the parameters acquired from different parts of the software supply chain that are based on the major components from the trust management domain such as reputation, experience, recommendation, penalty, etc. In other words, the computation of risks associated with software and the subsequent decision-making processes are approached and analyzed through the lens of trust. Furthermore, it accounts for context,
 integrates historical occurrences, and leverages timeliness to manage risk propagation. Figure \ref{fig:tree_diag_sriqt} outlines the proposed SAFER framework. The proposed framework is expected to have practical implications, including but not limited to (i) informed decision making, organizations and individual users can make well-informed choices about which software to adopt, (ii)  reducing security incidents by assisting users in identifying low risk software, (iii) risk reduction for organizations by identifying potential risk, and (iv) reassuring adoption of best security practices by highlighting the susceptible areas of the software supply chain.

Accordingly, 
the research at hand focuses on the following main contributions: 

\begin{itemize}

    \item Envisaging a software risk assessment framework from the trust viewpoint, SAFER (Software Analysis Framework for Evaluating Risk), which has not been undertaken previously to the best of our knowledge.  

    \item Providing new insights into risk assessment, including the influence of human factors and parameters associated with individual actors in a software supply chain, parameters besides code characteristics, software dependencies, context, historical behavior with relevant time-aware impact, recommendations or opinions, and penalty. The majority of these have not been considered by the state of the art in software risk management.

    \item Introducing novel non-subjective attributes (\emph{i.e.}, attributes that can be measured or assessed without personal bias or opinions) and quantifying these attributes and individual risk segments to yield consistent and non-subjective risk scores which the existing tools such as the Open Web Application Security Project (OWASP) risk calculator\cite{owasp} lack. Please note that subjectivity here does not refer to context-based subjectivity but bias or partiality.

    \item Employing suitable and rational influencing parameters as weights and catering to up-to-date/current risk factors while formulating contributing parameters to address the problem of weight quantification and achieving data-driven and dynamic weights associated with individual attributes and risk segments. This feature is not available in the existing tools such as the Open Source Security Foundation (OpenSSF) scorecard \cite{ossf} and OWASP risk calculator.

    \item Conducting an extensive evaluation, which demonstrates benefits over prior work, including mitigation of subjectivity and inconsistency, the impact of the human aspect, and data-driven dynamic computations.

\end{itemize}

The rest of the paper is organized as follows. Section~\ref{sec:literaturereview} reviews the 
state-of-the-art risk assessment models and tools. Section~\ref{sec:problemsetup} presents the problem statement, provides definitions for the major components, and specifies participating actors, Section~\ref{sec:ParameterSelection} examines the step-wise rationale behind the selected influencing parameters, Section~\ref{sec:proposedsystem} covers 
the details of our proposed SAFER's system model, whereas Section~\ref{sec:parametertuning} discuses the parameter tuning for tackling particular scenarios and tailoring specific elements within the envisaged SAFER framework. Section \ref{sec:results_analysis} offers insights into the dataset that has been employed and reports the results, and Section \ref{sec:discussion} presents a discussion regarding the claimed contributions, yielded results, and comparisons. Finally, Section~\ref{sec:conclusion} concludes the paper and outlines future directions.

\begin{table*}[!t]
\centering
\caption{Comparison with Existing Risk Assessment Tools.}%
\label{tab:comparison_sota}
{\renewcommand{\arraystretch}{0.5}
\scriptsize{
\begin{tabular}{ccccccccc}
\toprule
\textbf{Framework} & \textbf{Quantitative} & \textbf{Consistent}
& \textbf{Actor Element} & \textbf{Dynamic Weight} & \textbf{Data Driven Weight} & \textbf{Reputation} & \textbf{Context} & \textbf{Trust Perspective}\\
\tabularnewline
\midrule

OWASP \cite{owasp} & \xmark & \xmark & \xmark & \xmark & \xmark & \xmark & \xmark & \xmark\\\tabularnewline

OpenSSF Scorecard \cite{ossf} & \cmark & \cmark & \xmark & \xmark & \xmark & \xmark & \xmark & \xmark\\\tabularnewline

CVSS \cite{nistcvss} & \cmark & \xmark & \xmark & \xmark & \xmark & \xmark & \xmark & \xmark\\\tabularnewline

Risk Explorer \cite{riskexplorer} & \xmark & \cmark & \xmark & -- & -- & -- & -- & \xmark\\\tabularnewline

EU-Risk \cite{EUrisk} & \xmark & -- & \xmark & \xmark & \xmark & \xmark & \xmark & \xmark\\\tabularnewline
\textbf{SAFER} & \cmark & \cmark & \cmark & \cmark & \cmark & \cmark & \cmark & \cmark\\\tabularnewline
\bottomrule
\end{tabular}
}}
\end{table*}

\section{Related Work}
\label{sec:literaturereview}
This section presents a detailed literature review of the existing risk assessment works in both academia and the industry, followed by an overview of the contrasting features among the current tools (\emph{i.e.}, OpenSSF scorecard, and OWASP) and the proposed SAFER framework. Table \ref{tab:comparison_sota} exhibits a comparison between existing risk assessment tools in the industry and the proposed SAFER framework.  The calculators and risk evaluation methods mentioned were employed to assess risk, and the underlying literature on their design/development and functionality was thoroughly studied to deduce the conclusions presented in Table~\ref{tab:comparison_sota}.

\subsection{Literature Review}
Owing to the reliance of the digital industry, specifically software development and distribution, on open access and open source elements, adversaries are leveraging the existence of trust relationships among various stakeholders \cite{new62023}. To tackle these concerns, several frameworks have been presented, and several projects have been initiated, including but not limited to the Common Vulnerability Scoring System (CVSS) by the National Institute of Standards and Technology (NIST) \cite{nistcvss}, OWASP \cite{owasp}, OpenSSF scorecard \cite{ossf}, MITRE \cite{mitre}, Supply-chain Levels for Software Artifacts (SLSA) \cite{slsa}, EU-risk assessment \cite{EUrisk}, and Risk explorer \cite{riskexplorer}. 

CVSS by NIST established a publicly accessible framework to indicate the attributes and repercussions of different vulnerabilities related to IT systems. The national vulnerability database furnishes precise CVSS scores for known vulnerabilities \cite{nistcvss}. However, it brings in an element of subjectivity; it is difficult for us 
(1) to find how the scores aligned with severity are computed (\emph{e.g.}, target distribution low, medium, and high mapping on ranges 1\%-25\%, 26\%-75\%, and 76\%-100\%, respectively), and 
(2) to understand the reasoning behind adjustment or weighting factors (\emph{e.g.}, the base equation for impact has a factor of 10.41, or exploitability has a factor of 20, which has not been explained).

The OWASP risk rating methodology utilizes a conventional risk assessment approach, wherein risk is determined by both the probability of an attack occurring and the potential impact of a successful attack \cite{owasp}. However, it introduces subjectivity, employs pre-defined scores to qualitative factors without a data-driven approach, which is difficult to follow (\emph{e.g.}, score mappings of 1, 4, and 9 for motives), the overall likelihood and impact scores have been computed by averaging out all the scores for individual factors which implies that all the considered factors influence the final risk score equally. 

The OpenSSF scorecard by the Open Source Security Foundation is an automated tool designed to assess the security status of packages by analyzing the source code hosted on GitHub \cite{ossf, newossf}. However, it is unclear to us how the scores for individual parameters are decided (\emph{e.g.}, for the branch protection parameter in the OpenSSF scorecard, how are the scores 3, 6, 8, 9, and 10 decided for different tiers?). Moreover, the quantification of the degree of impact associated with the individual parameters is missing, \emph{i.e.}, the parameter dangerous overflow is assigned a severity level of critical, and a pre-defined static value for this weight has been assigned.  

MITRE presents a comprehensive framework (end-to-end) that prioritizes the firm integrity of the software supply chain across all stages of design, development, and delivery \cite{mitre}. Nonetheless, this encompasses solely software bill of materials (SBOM) derived metadata, infrastructure for software development, signing, and distribution.

Similarly, SLSA is SBOM-based, \emph{i.e.}, it offers transparency regarding the software components and ensures the integrity of SBOM. For open-source software, it guarantees to map all packages to the authorized list of packages and components. In contrast, for vendor-based software, the evaluation of the credibility of the assertions put forth by the vendors is emphasized due to the absence of source repositories \cite{slsa}. Nevertheless, the focus is solely on the building aspect; the trustworthiness of dependencies is not considered, consequently overlooking a thorough analysis of all the elements contributing to the software along with the code quality.

EU-risk assessment report discusses the escalated risk tied to software development and updating processes, vulnerabilities introduced due to misconfiguration, relying on a single supplier, etc. \cite{EUrisk}. However, it solely concentrates on the software development process and does not consider actor-based parameters, which suggests that it may not fully realize the entire scope of the threat landscape by overlooking the human factor. 

Risk explorer for software supply chains exhibits an attack tree and methodologies for malevolent code injection. Moreover, defenses against various attack techniques are also presented \cite{riskexplorer}. Nonetheless, it does not include quantitative analysis and scoring mechanisms, which makes it challenging to judge the risk severity and prioritize risks based on severity to make objective decisions. 

There is little research within the academic community that emphasizes risk assessment of software in software supply chain 
\cite{hyunchul, chess, donald, kumar, shaukat, pemmada}.

HyunChul et al. \cite{hyunchul} present risk as the combination of the likelihood of the vulnerability being in a specific state and the subsequent consequences. However, while defining the vulnerability risk index, the authors assume the vulnerabilities are statically independent. Moreover, the notion of vulnerability life cycle stages has been employed along with CVSS metrics, which brings subjectivity, potentially leading to inconsistent assessments. Brian Chess in \cite{chess} focuses primarily on the source code analysis only for risk assessment. 
Similarly, code complexity data and neural network-based risk analysis have been suggested in \cite{donald}. The authors take into consideration only the source code and sample-related metrics. 
In \cite{kumar}, the authors employ Bayesian belief networks for software risk assessment; however, like many others, they only rely on software development-related qualitative risk factors. 
Authors in \cite{shaukat} have proposed a dataset; however, they exclusively depend on software requirement risks and attributes drawn from experts and literature. 
Similarly, in \cite{pemmada}, only software requirement risk has been considered. 
The focus of academic research in risk management has primarily revolved around the project management viewpoint, which is mainly centered on risk assessment in software development projects.

Within the existing literature, there are four key limitations. First, the central part of software supply chain risk assessment research primarily focuses on qualitative analyses. Within quantitative risk evaluation, the quantification of weights has been overlooked, and instead, random static values of weights have been manually pre-defined. Second, there's a recurring absence of guidance regarding incorporating pre-defined adjustment factors into the formulation of such assessment models, which leads to subjectivity, inconsistency, and a lack of a data-driven dynamic approach to decision making. Third, a substantial proportion of these risk assessment frameworks fixates solely on the software development phase, thereby incorporating parameters solely concerning the software code itself. Moreover, the exclusion of software dependencies from risk management models constitutes a noteworthy limitation that may undermine the efficacy of these models by ignoring the security vulnerabilities and risks that may propagate through these dependencies. 
Lastly, the current research studies predominantly rely on conventional factors within the risk evaluation process, often neglecting the potential influence of the human elements (\emph{e.g.}, developer's experience, publisher's reputation, and user rating history) that significantly impact the threat landscape of the overall risk environment.


Accordingly, there is a compelling requirement for a risk assessment framework that effectively evaluates software within a software supply chain. This framework should encompass essential influencing parameters, including human factors, mitigate subjectivity, account for context, and leverage timeliness to manage risk propagation, 
and the integration of historical occurrences.

\begin{table*}[!t]
\centering
\caption{Comparison - SAFER with OWASP Risk Calculator \cite{owaspcalc} and OpenSSF Scorecard \cite{ossfcalc}.}%
\label{tab:comparison_tools}
{\renewcommand{\arraystretch}{0.50}
\scriptsize{
\begin{tabular}{p{1.25cm}|p{5.2cm}p{4.6cm}p{5.1cm}}
\toprule
\textbf{Features} & \textbf{OWASP} & \textbf{OpenSSF}
& \textbf{SAFER} \\
\tabularnewline
\midrule

\textbf{Quantitative} & Utilizes qualitative parameters only &  Utilizes a combination of qualitative and quantitative parameters & Represents both qualitative and quantitative attributes in terms of quantitative factors \\
\hline
\\
\textbf{Technical Parameters} & Focuses on qualitative attributes, which includes both technical and non-technical parameters & Considers technical parameters or risk factors only 
& Focuses on quantitative attributes, which includes both technical and non-technical parameters \\
\hline
\\
\textbf{Data Driven} & Individual checks and the weights assigned to each check are pre-defined static values & Individual checks and the weights assigned to each check are pre-defined static values & Relies solely on data-driven attribute checks as well as weights associated with individual parameters\\
\hline
\\
\textbf{Open Source} & Can be used for any project &  Focuses only on the open source projects on Github & With minimal adjustment, can be applied to any software application\\
\hline
\\
\textbf{Dynamic Weights} & Computes overall likelihood and impact scores by averaging all scores for individual factors, \emph{i.e.}, each parameter has an equal static and pre-defined weight assigned, implying that each considered factor equally influences the final risk score &  Employs static pre-defined weights while aggregating the overall score & Utilizes dynamic data-driven weights associated with individual parameters as well as the distinct risk segments\\
\hline
\\
\textbf{Inconsistency} & Evaluations for the same piece of software performed by different assessors in the same setting and context yield different results, which presents subjectivity and inconsistencies in the assessments & Yields consistent results & Utilizes quantifiable and objective parameters to produce consistent results\\
\hline
\\
\textbf{Computed Final Score} & Results yielded for the same piece of software computed by different assessors in the same setting and context are highly subjective and inconsistent & Comparison results suggest that there is a need to have additional features highlighting the human aspect, reputation and recommendations & Results are non-subjective, consistent and data-driven\\

\bottomrule
\end{tabular}
}}
\end{table*}
\subsection{Existing Tools}
\label{sec:existing tools}

We compare the proposed SAFER framework with the existing security/risk calculation tools OpenSSF scorecard and OWASP risk calculator by highlighting the differences and further explaining the need for the envisaged SAFER framework. Refer to Table~\ref{tab:comparison_tools} for details.

\section{Problem setup}
\label{sec:problemsetup}
If user \textbf{\textit{A}} needs to download a piece of software, \textbf{\textit{A}} should be able to assess if the software can be trusted by having a risk score computed for this software. As this risk evaluation is from the trust perspective, it is imperative for this evaluation to derive major components, \emph{e.g.}, context, reputation, and recommendations from trust management domain \cite{trust2023} for \textbf{\textit{A}} to have a holistic view. Risk is generally considered a multifarious abstraction that relies entirely on the subject's perception. According to NIST, risk refers to the probability or likelihood of a negative event or outcome occurring, often resulting in harm, loss, damage, or some form of adverse impact \cite{new8nistdef}. It involves situations where uncertainty exists about the future outcome, encompassing both the potential for something detrimental to happen and the magnitude of its consequences \cite{nist}.  
As mentioned earlier, trust and risk are closely linked. Evaluating the trustworthiness of an entity assists in realizing the potential risks associated with that entity.

\begin{definition} [\textbf{Trust}]
    According to ITU-T, trust is the quantifiable confidence and/or belief that reflects the aggregated value from past experiences and the anticipated value for the future \cite{ITU-Ttrust2022}.
\end{definition}

\begin{definition}[\textbf{Risk}]
    According to ISO 31000, risk is defined as the impact of uncertainty on achieving objectives. Here, the objectives refer to software security. It can be described as a combination of the potential consequences of an event and the likelihood of it occurring \cite{ISO_31000-Risk}. The higher the risk associated with a target entity, the higher the distrust of an entity towards that target entity becomes.
\end{definition}

As we are viewing risk with a trust perspective in our approach, the risk is defined, and subsequently, the risk assessment model is formulated using key components derived from the domain of trust.

\begin{definition} [\textbf{Context}]
    Context in the realm of software refers to the category or type of software being considered. This encompasses software designed for tasks like system automation or management, software aimed at enhancing security, or application software intended for diverse functionalities. The context of software influences the considerations and preferences when prioritizing risks \cite{trust2023}.
\end{definition}

\begin{definition} [\textbf{Reputation}]
    Reputation refers to the historical conduct of the entities involved in the software supply chain, including but not limited to software developers, publishers, and consumers. Among other parameters, the reputation is driven by the credibility demonstrated by these individual entities over a period of time \cite{acm2008}. 
\end{definition}

\begin{definition} [\textbf{Recommendations}]
    Recommendations refer to the opinions or ratings provided by users in relation to the target entity \cite{kang2019}. 
\end{definition}


In order to facilitate user \textbf{\textit{A}} in making a decision to opt for better and less risky software, quantification of the risk associated with the available software options is necessary. Quantifying risk relates to computing numerical values to diverse aspects of risk. This requires transforming qualitative data about risk into quantitative measures, allowing for a more precise and objective assessment of their potential effects. Through quantifying risk, individuals and organizations are able to assess and prioritize it in the order of significance or severity and make well-informed decisions regarding their mitigation.

The human elements, \emph{i.e.}, the actor-related characteristics such as the developer's experience, publisher's reputation, and user rating history, provide a comprehensive understanding of the threat landscape for the risk evaluation. Three key actors,  the developer, the publisher, and the user, play distinct yet interconnected roles within the software ecosystem, and their interactions with the software or parts of it significantly impact the overall dynamics of the supply chain. 

\subsubsection{Developer}
The developer, playing a central role among the key participants, is crucial for shaping the software through involvement in its design, coding, and comprehensive development. The developer's expertise and decision-making directly impact the quality, functionality, and security of the software product throughout the development phase.

\subsubsection{Publisher}
The publisher, being another vital contributor, assumes the role of distributing and ensuring the accessibility of the software to end users. This encompasses activities like packaging, marketing, and delivering the software through diverse channels. The actions taken by the publisher shape the market perception of the software and impact its accessibility to users.

\subsubsection{User}
The user, representing the final endpoint of the software supply chain, plays a vital role in the adoption, utilization, and feedback loop. Users interact with the software, provide insights, report issues, and contribute to the product's overall lifecycle. Their credibility, in turn, shapes the future development and iterations of the software.

Accordingly, a risk assessment model is required to assess risk scores associated with software within a supply chain. Since trustworthiness helps in understanding and realizing the corresponding risk, it is beneficial to model risk from the viewpoint of trust. Moreover, to include the human element, a comprehensive examination of the software supply chain, encompassing three pivotal actors: the developer, the publisher, and the user, is needed.

\section{Parameter Selection}
\label{sec:ParameterSelection}
Selecting the appropriate parameters for risk assessment of software from a trust perspective is a critical step in devising a trust-driven risk evaluation framework for ensuring the security of software systems. It is imperative to consider a multitude of factors that impact trust and security when determining the parameters that contribute to this assessment. For instance, (i) the parameters should be quantifiable, which means that each parameter should easily be expressed in numeric values; (ii) the parameters must be non-subject, which means that their numeric value can be measured objectively without any personal bias or opinions; (iii) the statistics of the software should be available (publicly available for the evaluation of open source software) which means that the values for each parameter must be accessible to the evaluator; (iv) the parameters should be simple for user-friendly calculation process, and (v) parameters like cost of software are not considered as cost does not impact the suitability of a piece of software as a less or more risky option. The process of cataloging various impacting parameters encompasses a blend of standards' requirements, published peer-reviewed work, expert consultation, and observable attributes associated with the software. Moreover, a structured evaluation of suitability attributes has been performed by employing the logic scoring of preferences (LSP) approach \cite{lsp2018}. 

\subsection{Step~1: Standards' Requirements}
We consider trust-influencing metrics while determining the parameters contributing to risk assessment. As mentioned in Section \ref{sec:problemsetup}, the ITU-T's definition of trust indicates that trust is partly dependent on past experiences~\cite{ITU-Ttrust2022}, which refers to reputation and recommendation. 

\subsection{Step~2: Human Element}
The involvement of the human element in the software supply chain is a critical aspect affecting the trust assessment parameters  \cite{newhumantrust2022, newhumantrust2023}, consequently impacting risk evaluation. The reputation, reliability, and experience of these actors significantly influence the trust placed in the software.

\subsection{Step~3: Published Peer Reviewed Work and Expert Opinion}
The published peer-reviewed work and the experts' opinions in the trust management domain suggest a significant influence of reputation, recommendation, context, penalty,
etc. \cite{sagar2022, trust2023, newtrust2024}. 

\subsection{Step~4: Observable Attributes}
It is essential to identify potential attributes present in the available data \cite{newobservable2019} to understand the information at hand and extract meaningful insights or conduct further analysis by employing this data. The parameters from (i) the dataset regarding \textit{top npm GitHub repositories}\footnote{https://www.kaggle.com/datasets/mikelanciano/top-npm-github-repositories}, (ii) GitHub, and (iii) Node Package Manager (NPM) have been utilized while devising the proposed risk analysis framework. Bhandari et al. \cite{bhandari} highlight the significance of code characteristics, including code length and the choice of programming language, in determining the software's security stance. Code length impacts the potential for vulnerabilities, with longer codebases potentially having more areas for exploitation. The programming language also introduces specific security challenges or advantages, which must be factored into the assessment. This highlights the importance of both quantitative and qualitative aspects of code. Another key consideration is that available observable attributes and data are not limited to the code. While code-related attributes like code complexity and language choice are vital, other factors play a crucial role in predicting vulnerabilities and identifying patterns of vulnerabilities. These additional attributes include the category of the software, the involved actors, known vulnerabilities, recommendations, and historical behaviors, among other attributes \cite{trust2023, newhumantrust2022, newhumantrust2023, ISO2018, ITU-Ttrust2022}. A comprehensive risk assessment considers these broader aspects to gain a holistic view of the software's trustworthiness.

\subsection{Step~5: Logic Scoring of Preferences}
Logic scoring of preferences (LSP) has been employed as a structured approach to evaluate suitability attributes. The primary objective of such methods is to offer a quantitative approach for making evaluation decisions based on criteria that accurately represent stakeholder goals and requirements. LSP works on the fundamental notion of completely aligning with human intuitive processes of evaluation and selection \cite{lsp2018}. The suitability attributes for assessing the software security in the proposed framework do not include any cost-related attributes, as the cost does not impact the suitability of a piece of software as a less or more risky option. The stepwise decomposition method to develop the suitability attribute tree starts with the identification of four categories of attributes, including (i) reputation/experience, which contains parameters related to the experience and reputation of the actors (i.e., developer's experience in programming language, the developer's expertise, and the publisher's experience) (ii) exposure, which includes parameters related to susceptibility (i.e., dependencies, code coverage, unresolved vulnerabilities, and resolved vulnerabilities), (iii) popularity and maintenance, which covers parameters that describe the ongoing development and acceptance of software  (i.e., frequency updates, rating, and downloads), and (iv) supplements, which encompasses attributes whose changing value will not enhance nor weaken the security of software, however, these can help in the assessment of the security of a software in conjunction of other parameters (i.e., code length, forks, and context). 

In essence, by carefully selecting and analyzing these parameters, it becomes possible to build a comprehensive and effective risk assessment framework that enhances the security and trustworthiness of software in a software supply chain. This multifaceted approach is vital for addressing the ever-evolving landscape of software vulnerabilities and threats, ultimately ensuring the reliability of software systems.

\section{System Model}
\label{sec:proposedsystem}

\noindent Let an individual software be $S_i$, wherein $i \in \{1, 2, 3, ..., I\}$, and a time instance is denoted by $t_k$, where $k \in \{1, 2, 3, ..., T\}$. Please note that the time index is being used to acknowledge the dynamic nature of risks, which can change with evolving circumstances or influencing risk factors. It is crucial to know when the risk computation took place, or in other words, how old the risk assessment is. 
As mentioned in Section \ref{sec:ParameterSelection}, the proposed risk assessment framework employs LSP for the suitability attributes' aggregation. The overall risk score has been divided into three risk segments corresponding to each actor. Within these risk segments, the LSP concept of maximum \textit{substitutability} or \textit{orness} has been utilized, and all the degrees of orness/andness have been decided by data-driven weights. Moreover, all suitability attributes are objectively quantifiable, independent, and can be measured individually. In addition, it is noteworthy that the notion of substitutability also brings in the property of compensativeness, which helps accommodate the unavailability of data against any parameter(s). Furthermore, if a risk segment cannot be evaluated due to limited or no data availability, or due to an undefined output, it must be assumed that this particular risk segment carries the maximum possible risk (i.e., normalized value equal to 1). In case there is data available only for some of the parameters within a risk segment data, the remaining parameters can be put as zero while computing that risk segment. 


\subsection{Developer based Risk
:}

The developer-based risk ($R^{\textup{DEV}}_{S_i,t_k}$) is an amalgamation of the risk from code dependencies, code specifications, programming language, historical information on vulnerabilities, and the developer's experience associated with the software $S_i$ at time instance $t_k$. The LSP property of compensativeness applies to all three risk segments that fall under the developer-based risk.

\subsubsection{Code Dependencies
}
The code dependencies-based risk ($R^{\textup{CD}}_{S_i,t_k}$) depends on the dependencies of the software $S_i$ at time instance $t_k$. Please note that dependencies here (in this manuscript) refer to other 3rd party software and packages rather than libraries\footnote{Software package comprises a set of modules, whereas a software library consists of pre-written code.}. 

\noindent \textit{Motivation:} Literature and security solution providers indicate that a higher number of dependencies increases risk and suggest minimizing dependencies to mitigate associated risks \cite{newdep1, newdep2}. It is worth noting that to reduce subjectivity, factors such as the quality of dependencies have been omitted from consideration as suitability attributes while applying the LSP process.

We have
\begin{equation}
R^{\textup{CD}}_{S_i,t_k}= D_{S_i,t_k},
  \label{eq:riskcodedependencysimple}
\end{equation}
where $R^{\textup{CD}}_{S_i,t_k}$ is the risk from code dependencies of the software, and $D_{S_i,t_k}$ is the number of dependencies for the software $S_i$ at time instance $t_k$. 

\subsubsection{Code Specifications
}
The code specifications-based risk is a combination of code characteristics, \emph{i.e.}, length, characteristics, and the relevant expertise of the developer for software $S_i$ at time $t_k$. 

\noindent \textit{Motivation:} Literature suggests the reliance of software risk on the code characteristics \cite{donald,newcodechrctr1}. Moreover, studies indicate the impact of the developer's reputation on different phases of the software review process \cite{newdevrep1}. Multiplication operation is employed to capture the proportion contribution of the code characteristics and to intensify the effect of vulnerabilities in $R^{\textup{CS}}_{S_i,t_k}$. As the code length belongs to the supplement category of suitability attributes in LSP, the property of simultaneity has been applied here.

We have
\begin{equation}
\label{eq:riskcodespecsimple}
      R^{\textup{CS}}_{S_i,t_k} = w^{\textup{LC}}_{t_k}.l_{S_i,t_k},
\end{equation}
where $R^{\textup{CS}}_{S_i,t_k}$ is the risk from code specifications, $w^{\textup{LC}}_{t_k}$ is the weight associated with the code length based on the historical reputation of the developer of software $S_i$ at time $t_k$, and $l^{\textup{c}}_{S_i}$ is the total number of lines in the code. 

We have

\begin{equation}
\label{eq:weightcodelength}
   w^{\textup{LC}}_{t_k}=  \frac{\sum_{j}\sum_{i=1}^{I}V^{\textup{D}}_{j,S_i,t_k}}{\sum_{j}|S^{\textup{D}}_{j,t_k}|},
\end{equation}
where $V^{\textup{D}}_{S_i,t_k}$ is the number of vulnerabilities in all the software developed by the same developer, $j$ denotes a developer, and $S^{\textup{D}}_{t_k}$ is the set of all software developed by the same developer at time $t_k$. Simply put, the weight $w^{\textup{LC}}_{t_k}$ represents the average number of vulnerabilities per software developed by a developer. It is worth noting that for the developer who has previously developed software with no vulnerabilities, the weight $w^{\textup{LC}}_{t_k}$ will be zero regardless of the total number of software developed. In other words, even if the developer has developed only a couple of software before and no vulnerability has been reported on that software, the developer will be given a chance with the code specification-based risk.

\subsubsection{Programming Language
}
The programming language-related risk ($R^{\textup{PL}}_{S_i,t_k}$) relies on the proportion of the experience of the developer(s) in terms of the number of years in a programming language to the overall experience in software development.

\noindent \textit{Motivation:} Literature indicates that extensive experience within a domain is essential to achieve exceptional levels of performance \cite{newexprncprfrm1}. In other words, trust and distrust towards an entity rely on the experience. The property of substitutability has been employed and 
the subtraction from 1 indicates the inverse effect, \emph{i.e.}, as the experience proportion increases, the risk associated with it decreases.

We have
\begin{equation}
\label{eq:riskpub}
R^{\textup{PL}}_{S_i,t_k}=\begin{cases}
    1-\frac{\sum_{j} E^{\textup{DS}}_{j,S_i,t_k}}{\sum_{j} E^{\textup{DA}}_{j,S_i,t_k}}, & \textit{if $\sum_{j}E^{\textup{DA}}_{j,S_i,t_k} > 0$},\\
    1, & \textit{if $\sum_{j}E^{\textup{DA}}_{j,S_i,t_k} = 0$},
\end{cases}
\end{equation}
where $R^{\textup{PL}}_{S_i,t_k}$ is the risk from the programming language, $j$ denotes a developer, $E^{\textup{DS}}_{j,S_i,t_k}$ represents the experience of a developer in terms of the number of years in the same language, whereas $E^{\textup{DA}}_{j,S_i,t_k}$ is the total number of years that developer has been developing software. 

\subsubsection{Aggregation of A.1, A.2, and A.3}
To compute the developer's overall risk ($R^{\textup{DEV}}_{S_i,t_k}$), the 
individual risk segments \emph{i.e.}, the risks related to the code dependencies ($R^{\textup{CD}}_{S_i,t_k}$), code specifications ($R^{\textup{CS}}_{S_i,t_k}$), and programming language ($R^{\textup{PL}}_{S_i,t_k}$) have been aggregated by applying logic neutrality in LSP process, i.e., the weighted sum of individual risk segments as: 
\begin{equation}
\label{eq:totaldeveloperrisksimple}
R^{\textup{DEV}}_{S_i,t_k}=w^{\textup{CD}}_{S_i,t_k}.R^{\textup{CD}}_{S_i,t_k} + w^{\textup{CS}}_{S_i,t_k}.R^{\textup{CS}}_{S_i,t_k} + w^{\textup{PL}}_{S_i,t_k}.R^{\textup{PL}}_{S_i,t_k},
\end{equation}
where $R^{\textup{DEV}}_{S_i,t_k}$ is the total risk associated with the developer, $R^{\textup{CD}}_{S_i,t_k}$ is the risk linked to the code dependencies and $w^{\textup{CD}}_{S_i,t_k}$ is the weight associated with it, $R^{\textup{CS}}_{S_i,t_k}$ is the risk connected with the code specifications and $w^{\textup{CS}}_{S_i,t_k}$ is the weight for it, whereas $R^{\textup{PL}}_{S_i,t_k}$ is the risk related to the programming language and $w^{\textup{PL}}_{S_i,t_k}$ is the weight linked with it for software $S_i$ at time $t_k$.

\noindent \textit{Motivation:} 
Code coverage quantifies how thoroughly a test suite tests a software system \cite{codecvrgdef}. Literature suggests that code coverage helps in maintenance tasks like predicting the likelihood of faults in a software \cite{newcodecvrg1, newcodecvrg2}. The subtraction from 1 indicates the inverse effect, i.e., as the code coverage increases, the impact of the associated risk decreases.

We have
\begin{equation}
\label{eq:weightforcodedependencyrisk}
w^{\textup{CD}}_{S_i,t_k} = 1-C^{\textup{C}}_{S_i,t_k},
\end{equation}
where $w^{\textup{CD}}_{S_i,t_k}$ is the weight associated with the code dependency risk, and $C^{\textup{C}}_{S_i,t_k}$ is the code coverage.

\noindent \textit{Motivation:} 
As mentioned above, the expertise and experience of the developer play a significant role in developing high-performing software \cite{newexprncprfrm1}. An exponential function with a negative power is used to emphasize the smaller positive values inversely. Moreover, compared to alternative functions such as $\tanh$, it maps all real numbers to positive real numbers and is more straightforward to scale its value between 0 and 1.

We have
\begin{equation}
w^{\textup{CS}}_{S_i,t_k}= e^{- E^{\textup{DB}}_{S_i,t_k}}, 
  \label{eq:weightforcodespec}
\end{equation}
where $E^{\textup{DB}}_{S_i,t_k}$ is the expertise/experience of the developer in terms of programming language at time $t_k$, and its value lies in the interval [0, 1];  

\begin{equation}
\label{eq:Ed2}
E^{\textup{DB}}_{S_i,t_k}=\frac{\sum_{j}|S^{\textup{L}}_{j,t_k}|}{\sum_{j}|S^{\textup{D}}_{j,t_k}|},
\end{equation}
%
%
where $S^{\textup{L}}_{t_k}$ is the set of all previously developed software by the same developer in the same language, $j$ denotes a developer, and $S^{\textup{D}}_{t_k}$ is the set of all previously developed software by the same developer. 

We have
\begin{equation}
\label{eq:devweightsum}
   w^{\textup{PL}}_{S_i,t_k} = |1 - (w^{\textup{CD}}_{S_i,t_k} + w^{\textup{CS}}_{S_i,t_k})|.
\end{equation}

\noindent \textit{Note:} The absolute sum of all weights is calibrated to be equal to 1. Literature suggests that this is an established practice \cite{newtranits2023}.

\subsection{Publisher based Risk
:}
The risk associated with a publisher of software in a software supply chain is defined as the amalgamation of the update frequency and the reputation/experience of the said publisher. 

\noindent \textit{Assumption:}
Similar to the developer's experience, the experience and expertise of all actors involved impact the risk associated with software. The average number of software published per year and the frequency of updates required for that software are inversely related to the risk segment. The property of substitutability has been utilized while applying the LSP process.  

We have
\begin{equation}
\label{eq:modifiedpubrisksimple}
R^{\textup{PB}}_{S_i,t_k}= \begin{cases}
    \frac{\sum_{x} E^{\textup{PY}}_{x,S_i,t_k}}{\sum_{x} E^{\textup{PX}}_{x,S_i,t_k}} . \frac{1}{F^{\textup{UP}}_{S_i,t_k}}, & \textit{if $E^{\textup{PX}}_{x,S_i,t_k}, F^{\textup{UP}}_{S_i,t_k} \neq 0$},\\
    1, & \textit{if $E^{\textup{PX}}_{x,S_i,t_k}, F^{\textup{UP}}_{S_i,t_k}=0$},
\end{cases}
\end{equation}
where $R^{\textup{PB}}_{S_i,t_k}$ is the risk associated with the publisher, $E^{\textup{PX}}_{S_i,t_k}$ is the experience of the publisher in terms of the number of software published, $x$ represents a publisher, $F^{\textup{UP}}_{S_i,t_k}$ is the update frequency of the software, and $E^{\textup{PY}}_{S_i,t_k}$ is the experience of the publisher in terms of years. Please note that the update frequency is in the range (0, 1]. 

\subsection{User indicated Risk
:}
The risk indication provided by a user regarding software is defined as the proportion of users who indicated satisfaction with the software compared to those who downloaded it.

\noindent \textit{Motivation:}
The observed distribution of data suggests that consumer trust in software is represented by the ratio of users who expressed satisfaction with the software to those who downloaded it. The property of substitutability has been applied, and the subtraction from 1 implies the inverse effect, \emph{i.e.}, as the satisfaction proportion increases, the risk associated with it decreases. 

We have
\begin{equation}
\label{eq:modifieduserrecommrisk}
R^{\textup{UR}}_{S_i,t_k} = 1- \frac{G^{\textup{US}}_{S_i,t_k}}{N^{\textup{DS}}_{S_i,t_k}},
\end{equation}
where $R^{\textup{UR}}_{S_i,t_k}$ is the risk associated with a particular software according to the recommendation or opinion from users, $G^{\textup{US}}_{S_i,t_k}$ is the rating (\emph{i.e.}, number of stars) assigned by the users to specific software, and $N^{\textup{DS}}_{S_i,t_k}$ is the total number of times that software has been downloaded. 

\subsection{Penalty $(0 \leq P_{S_i,t_k} \leq 1)$:}
Penalty ($P_{S_i,t_k}$) is the increment in the final risk score ($R^{\textup{F}}_{S_i,t_k}$) of a software $S_i$ at time instance $t_k$ and is associated with the software which has established a particular degree of mistrust. 

\noindent \textit{Motivation}:
Exponential expression is used to emphasize the smaller positive values. The subtraction from 1 specifies the vulnerabilities result in an inverse effect, \emph{i.e.}, as the exponential expression increases, the penalty associated with it decreases. It is noteworthy that for the same base value, a higher proportion of unfixed vulnerabilities results in a higher value, whereas if the base value is increased, the impact of the unresolved vulnerabilities is less pronounced. The context parameter is employed to adjust the stringency of penalties based on the software category, while the proportion of unresolved vulnerabilities reflects the level of susceptibility. Literature in the trust management domain suggests the use of pre-defined fixed values for penalty factors. However, in the proposed risk evaluation framework, we aim for a data-driven and dynamic penalty factor \cite{newpenal2021}. Context is taken as a mandatory input attribute to compute penalty while applying the LSP process. However, if the context is unknown, it is assumed that the software is security software for the most rigorous risk assessment.

We have
\begin{equation}
    P_{S_i,t_k} = 1-({C^{\textup{TXT}}_{S_i}})^{V^{\textup{UP}}_{S_i,t_k}},
  \label{eq:penalty}
    \end{equation}
where $P_{S_i,t_k}$ is the penalty factor, and $V^{\textup{UP}}_{S_i,t_k}$ is the proportion of unresolved vulnerabilities. Please note that if the software has no vulnerabilities, $V^{\textup{UP}}_{S_i,t_k} = 0$. ${C^{\textup{TXT}}_{S_i}}$ is the context of the software in question and
\begin{equation} 
C^{\textup{TXT}}_m = \begin{cases}
    0.2, & \text{if security software},\\
    0.3, & \text{if automation software},\\
    0.5, & \text{otherwise},
  \end{cases}
\label{eq:ctxt}
\end{equation}
where $C^{\textup{TXT}}_m$ is the context of the $m^{th}$ software being evaluated. In this paper, we consider $M=3$ cases of context, \emph{i.e.}, security software (\emph{i.e.}, $m=s$), automation software (\emph{i.e.}, $m=a$), and other software (\emph{i.e.}, $m=o$). \\
\noindent \textit{Note:} The values of 0.2, 0.3, and 0.5 selected for context are chosen by keeping the sum of all equal to 1. These values can be customized to align with the stringent requirements of the organization.

We have
\begin{equation} 
\sum_{m=1}^{M}C^{\textup{TXT}}_m = 1.
\label{eq:ctxtsummation}
\end{equation}

\subsection{Final Risk Score $(0 \leq R^{\textup{F}}_{S_i,t_k} \leq 1)$:}
The final risk score ($R^{\textup{F}}_{S_i,t_k}$) combines the actor (\emph{i.e.}, developer, publisher, and user) based risk components into a single score by applying logic neutrality in LSP process, i.e., the weighted sum of individual actor-based risk scores as $ w^{\textup{DEV}}_{S_i,t_k}.R^{\textup{DEV}}_{S_i,t_k} + w^{\textup{PB}}_{S_i,t_k}.R^{\textup{PB}}_{S_i,t_k} + w^{\textup{UR}}_{S_i,t_k}.R^{\textup{UR}}_{S_i,t_k}$. However, to enable comparisons among various risk scores, the risk has been depicted on a scale ranging from 0 to 1 by employing the sigmoid function \cite{newsigmoid}. This guarantees uniform interpretation for comparability. Moreover, to make the resulting score gradually increase, the sigmoid function has been modified, and the adjustment factors, \emph{e.g.}, 4 and 0.04, have been determined by graph analysis to analyze the potential outcomes. The resultant final risk $R^{\textup{F}}_{S_i,t_k}$ is depicted as:
\begin{equation}
\label{eq:finalrisk}
R^{\textup{F}}_{S_i,t_k} = \frac{1}{1+e^{4-0.04(w^{\textup{DEV}}_{S_i,t_k}.R^{\textup{DEV}}_{S_i,t_k} + w^{\textup{PB}}_{S_i,t_k}.R^{\textup{PB}}_{S_i,t_k} + w^{\textup{UR}}_{S_i,t_k}.R^{\textup{UR}}_{S_i,t_k})}},
\end{equation}
and
\begin{equation}
\label{eq:finalweightssum}
w^{\textup{DEV}}_{S_i,t_k} + w^{\textup{PB}}_{S_i,t_k} + w^{\textup{UR}}_{S_i,t_k} = 1.
\end{equation}

\noindent \textit{Note:} The sum of all weights is calibrated to be equal to 1.

\noindent \textit{Motivation:} Forks enable developers to conduct code reviews and audits by comparisons between the original codebase and modifications made in the forked repositories. This enables the identification of potential security vulnerabilities, compliance issues, or other risks, allowing collaborative efforts to address them effectively \cite{newgithubfork}. The higher the number of forks, the lesser the weight of the corresponding risk will be.

We have
\begin{equation}
\label{eq:userweight}
w^{\textup{DEV}}_{S_i,t_k} = \begin{cases}
    \frac{1}{f_{S_i,t_k}}, & \text{if $f_{S_i,t_k} > 0$},\\
    1, & \text{if $f_{S_i,t_k} = 0$},
  \end{cases}
\end{equation}
where $f_{S_i,t_k}$ is the number of forks given to the software.

\noindent \textit{Motivation:} The literature suggests that the probability of unresolved vulnerabilities can be computed by employing the following equation \cite{newbtcmv2022}. The higher the likelihood of unresolved vulnerability, the higher the weight of the risk will be.

We have
\begin{equation}
\label{eq:developerweight}
w^{\textup{PB}}_{S_i,t_k} = \frac{V^{\textup{U}}_{S_i,t_k} + 1}{V^{\textup{R}}_{S_i,t_k}+V^{\textup{U}}_{S_i,t_k}+2},
\end{equation}
where $V^{\textup{R}}_{S_i,t_k}$ are the number of known vulnerabilities resolved, and $V^{\textup{U}}_{S_i,t_k}$ are the number of known vulnerabilities unresolved.

\subsection{Final Risk Score with Penalty $(0\leq R^{\textup{FP}}_{S_i,t_k} \leq 1)$}
The final risk score with a penalty ($R^{\textup{FP}}_{S_i,t_k}$) is the risk score of a software $S_i$ at time $t_k$ after the penalty has been applied. This risk score 
is computed as:
\begin{equation}
\label{eq:riskafterpenalty}
R^{\textup{FP}}_{S_i,t_k} = \begin{cases}
   R^{\textup{F}}_{S_i,t_k} + P_{S_i,t_k}, & \text{if }
   \begin{aligned}[t]
       R^{\textup{F}}_{S_i,t_k} > 0.5, \\
       \textup{\& } R^{\textup{F}}_{S_i,t_k} + P_{S_i,t_k} < 1,
       \end{aligned}
   
   \\
    1, & \text{if $R^{\textup{F}}_{S_i,t_k} \geq (1-P_{S_i,t_k})$},\\
    R^{\textup{F}}_{S_i,t_k}, & \text{otherwise}.
  \end{cases}
\end{equation}

\subsection{Risk Bands}
The resulting risk scores are categorized into four bands, facilitating the communication and classification of the risk level linked to a specific software. Each band typically represents a range of scores corresponding to a particular level of risk, allowing for more straightforward interpretation and decision-making. The process works as follows:

\subsubsection{Low Risk Band}
$0 \leq R^{\textup{FP}}_{S_i,t_k} < 0.25$\\
This indicates that the risk associated with the assessed software is minimal. It suggests that the likelihood of negative outcomes is low, and the impact, if any, is expected to be minor.

\subsubsection{Moderate Risk Band} $0.25 \leq R^{\textup{FP}}_{S_i,t_k} < 0.5$\\
This suggests a moderate level of risk. While the likelihood of adverse events might be higher than in the low-risk band, the potential impact is still manageable. Attention and monitoring may be required to prevent escalation.

\subsubsection{High Risk Band} $0.5 \leq R^{\textup{FP}}_{S_i,t_k} < 0.75$\\
This signifies a significant level of risk. There is a higher likelihood of adverse events occurring, and the potential impact is substantial. Immediate action or intervention may be necessary to mitigate the risk and prevent severe consequences.

\subsubsection{Critical Risk Band} $0.75 \leq R^{\textup{FP}}_{S_i,t_k} \leq 1$\\
This represents the highest level of risk. The likelihood of severe consequences is extremely high, and immediate and decisive actions are typically required to address the situation. This band often signifies a potential crisis or emergency that demands urgent attention.

\begin{table*}[!b]
\begin{center}

\caption{Structure of the Dataset.}%
\label{tab:datastructure}
{\renewcommand{\arraystretch}{0.05}
\scriptsize	{
\begin{tabular}{p{0.5cm}p{0.40cm}p{0.3cm}p{0.5cm}p{0.6cm}p{0.6cm}p{0.61cm}p{0.55cm}p{0.55cm}p{0.55cm}p{0.55cm}p{0.55cm}p{0.55cm}p{0.55cm}p{0.55cm}p{0.5cm}p{0.5cm}}
\toprule
\textbf{\centering\rotatebox[origin=c]{90}{Sample}} &
\textbf{\centering\rotatebox[origin=c]{90}{Code Length}} & 
\textbf{\centering\rotatebox[origin=c]{90}{Developer}}  &
\textbf{\centering\rotatebox[origin=c]{90}{Publisher}}  &
\textbf{\centering\rotatebox[origin=c]{90}{Year}} & 
\textbf{\centering\rotatebox[origin=c]{90}{Language}} & \textbf{\centering\rotatebox[origin=c]{90}{Update Frequency}}&
\textbf{\centering\rotatebox[origin=c]{90}{Forks}}& 
\textbf{\centering\rotatebox[origin=c]{90}{Downloads}}& \textbf{\centering\rotatebox[origin=c]{90}{Unresolved Vulnerabilities}}& 
\textbf{\centering\rotatebox[origin=c]{90}{Known Vulnerabilities}}& 
\textbf{\centering\rotatebox[origin=c]{90}{Dependencies}}& 
\textbf{\centering\rotatebox[origin=c]{90}{Rating}}&
\textbf{\centering\rotatebox[origin=c]{90}{Code Coverage}}&
\textbf{\centering\rotatebox[origin=c]{90}{Context}}\\
\tabularnewline
\midrule
\centering{1} & \centering{222} & \centering{V} & \centering{L} & \centering{2019} & \centering{Python} & \centering{0.77} & \centering{562} &  \centering{15181} & \centering{0} & \centering{2294} & \centering{26} & \centering{5037} & \centering{0.97}& \centering{0.2} \\\tabularnewline

\centering{2} & \centering{272} & \centering{W} & \centering{H} & \centering{2017} & \centering{C} & \centering{0.35} & \centering{1961} &  \centering{27459} & \centering{5} & \centering{1576} & \centering{6} & \centering{15093} & \centering{0.85}& \centering{0.3} \\\tabularnewline[-0.2cm]

\centering{~\vdots} & \centering{} & \centering{} & \centering{} &  \centering{}& \centering{}& \centering{}& \centering{}& \centering{}& \centering{}& \centering{}& \centering{}& \centering{}& \centering{}&\centering{~\vdots}\\\tabularnewline

\centering{4500} & \centering{305} & \centering{H} & \centering{S} & \centering{2023} & \centering{Python} & \centering{0.81} & \centering{541} &  \centering{7605} & \centering{1} & \centering{48} & \centering{0} & \centering{1661} & \centering{0}& \centering{0.3} \\\tabularnewline

\centering{4501} & \centering{663} & \centering{N} & \centering{O} & \centering{2020} & \centering{Java} & \centering{0.93} & \centering{4443} &  \centering{80721} & \centering{799} & \centering{11909} & \centering{0} & \centering{37328} & \centering{0.91}& \centering{0.5} \\\tabularnewline[-0.2cm]

\centering{~\vdots} & \centering{} & \centering{} & \centering{} &  \centering{}& \centering{}& \centering{}& \centering{}& \centering{}& \centering{}& \centering{}& \centering{}& \centering{}& \centering{}&\centering{~\vdots}\\\tabularnewline

\centering{8999} & \centering{48} & \centering{R} & \centering{D} & \centering{2018} & \centering{Python} & \centering{0.004} & \centering{260} &  \centering{1733} & \centering{17} & \centering{868} & \centering{16} & \centering{216} & \centering{0.88}& \centering{0.2} \\\tabularnewline

\centering{9000} & \centering{138} & \centering{N} & \centering{J} & \centering{2017} & \centering{C} & \centering{0.86} & \centering{161} &  \centering{2579} & \centering{4} & \centering{289} & \centering{11} & \centering{559} & \centering{0}& \centering{0.2} \\\tabularnewline[-0.2cm]
\bottomrule
\end{tabular}
}}
\end{center}
\end{table*}

\section{Parameter Tuning}
\label{sec:parametertuning}

The envisaged SAFER framework is adaptable to suit the particular needs of the organization. Moreover, it is essential to address specific scenarios (\emph{e.g.}, when there exists no dependencies or the publisher is a new publisher) that may arise. This section elaborates on both the customization options and occurrences of particular situations that may require modification to the risk assessment process. Please note that some customization options can add subjectivity between organizations; however, it will keep the computations within the organization non-subjective and consistent as the tuning will represent the context and not partiality.

\subsection{Greater Dependencies' Influence}
\label{sec:dependecyimpact}
The influence of dependencies can be heightened by incorporating a sensitivity factor when computing $R^{\textup{CD}}_{S_i,t_k}$, which is defined by the organization to align with their security requirements, by multiplying it with $D_{S_i,t_k}$ in equation~\ref{eq:riskcodedependencysimple}.

\subsection{No Dependencies}
If the said software code does not have any external dependencies, the resulting risk based on code dependencies $R^{\textup{CD}}_{S_i,t_k}$ will be zero. 

\subsection{New Publisher}
If the publisher of the software is new, \emph{i.e.}, if they lack prior experience ($E^{\textup{PX}}_{S_i,t_k}$ and/or $E^{\textup{PY}}_{S_i,t_k}$ is zero), the resulting risk based on the publisher $R^{\textup{PB}}_{S_i,t_k}$ attains its maximum value of 1.

\subsection{Customized Risk Bands}
In risk management, a one-size-fits-all approach may not adequately address each organization's unique characteristics and security requirements. Organizations have the autonomy to customize their risk bands to align with their business objectives, reflect the evolving threat landscape, and adapt to industry regulations.

\subsection{Customized Weights}
The proposed risk assessment model employs dynamic, \emph{i.e.}, data-driven weights relying on risk-influencing parameters to adapt to the constantly evolving security considerations in the software supply chain. However, these weights can be tailored to align with the stringent requirements of the organization. Please note that customizing weights can add subjectivity between organizations; however, it will keep the computations within the organization consistent as the tuning will represent the context and not partiality.

\subsection{Calibrating or Normalizing Risk Score}
To maintain a risk score within the range of 0 to 1, it is necessary to calibrate or normalize the computed risk score. This practice is logical as it compares values against similar metrics used in other software or packages within the organization. Typically, organizations rely on a predefined set of approved or commonly used software to carry out their tasks. Assessing risk about other software within this set is a prudent approach, providing a meaningful context for evaluating potential risks associated with the organization's software ecosystem.

\subsection{Penalty}
The penalty threshold has been set by taking a mean of the minimum and maximum possible risk values, \emph{i.e.}, 0 and 1, respectively. This threshold is adjustable based on the stringent requirements of individual users/organizations to make the risk evaluation criteria more or less strict.
Moreover, the values of 0.2, 0.3, and 0.5 selected for context in equation~\ref{eq:ctxt} can be customized to align with the stringent requirements of the organization.

\section{Results}
\label{sec:results_analysis}

In this section, we provide insights into the dataset utilized for calculating risk through the proposed SAFER framework. Furthermore, yielded results are also presented.

\subsection{Dataset}
\label{sec:dataset}

The data employed to compute risk is taken from (i) the data regarding \textit{top npm GitHub repositories}, which is a combination of data taken from platforms for software development and software hosting, \emph{e.g.}, GitHub and NPM, and (ii) synthetic data for remaining attributes. 

While it is technically feasible to extract data for these remaining attributes from GitHub and NPM repositories, the process involves complex data mining and analysis techniques beyond this current study's scope. We have generated representative values for these parameters to demonstrate the efficacy of our proposed risk analysis framework without compromising its validity. This approach allows us to validate the framework's functionality while maintaining focus on its core methodological contributions. The criterion used to generate this synthetic data has been provided in the appendix.

Let $\mathcal{D}$ denote the entire dataset and $\mathcal{D}_{x} \subset \mathcal{D}$ denote the subset of the dataset, excluding the parameters and their corresponding values that were generated. Please note that $\mathcal{D}_{x}$ only includes code dependencies, code coverage, user ratings, forks, vulnerabilities resolved, and total vulnerabilities. Table \ref{tab:datastructure} provides the entire dataset's ($\mathcal{D}$) structure and a few samples. More details can be found in the Appendix.

\subsection{Risk Assessment}
\label{sec:assessment}
Table \ref{tab:computations} presents the final outcomes via the proposed SAFER framework for the selected samples. An example of how SAFER will be used in practice has been provided in the Appendix.   

\subsection{Evaluation \& Comparison}
\label{sec:evaluation}
The evaluation of the proposed SAFER framework has been performed based on user experience and usability tests. It is a well-established practice to conduct software evaluations through user experience. This approach ensures that the software meets usability standards and aligns with user needs and expectations \cite{UX1, UX2, UX3, UX4, UX5}. The user experience has been assessed via usability, satisfaction, efficiency, learnability, and accessibility. To remove subjectivity from the opinions, a binary scale has been used for each user experience attribute, i.e., 0 represents bad user experience, whereas 1 represents good user experience.

Comparisons for the user experience, as well as risk scores generated by the OpenSSF scorecard, OWASP risk calculator, and the proposed SAFER framework, have been provided to evaluate and validate the necessity of the parameters employed by the proposed SAFER framework. Moreover, tests related to errors and missing information against parameters have been performed, and assumptions are embedded in the functionality of the framework for added convenience. More details on these tests can be found in the Appendix.


Table \ref{tab:userexperience} outlines the comparison of user experiences associated with the OpenSSF scorecard, OWASP risk calculator, and the proposed SAFER framework for risk assessment of private repositories. 
Table \ref{tab:comparisonossf} exhibits the comparison of scores computed by OpenSSF scorecard \cite{ossfcalc} and the proposed SAFER framework along with the manual verification. Moreover, Figures \ref{fig:sensitivityheatmaps} and \ref{fig:dynamicityheatmaps} demonstrate the sensitivity and the dynamic aspect of the proposed SAFER framework in the form of heat maps, respectively. Furthermore, Table \ref{tab:comparisonwwo} illustrates the comparison of the risk scores calculated using the proposed SAFER framework by first including and then excluding the parameters for which self-generated data points were utilized, whereas Table \ref{tab:riskowasp5} delineates the risk scores computed by 5 assessors using OWASP risk calculator \cite{owaspcalc}. 

\section{Discussion}
\label{sec:discussion}
This section explores how well the proposed SAFER framework aligns with its stated contributions in Section \ref{sec:intro} followed by a discussion on computed outputs.

\subsection{Contributions}
As mentioned earlier, the envisaged SAFER framework aimed to (i) quantify risk to eliminate subjectivity and inconsistent risk assessments, (ii) explore the human element, \emph{i.e.}, the risk associated with the actors engaged in the software supply chain, (iii) employ data-driven and dynamic weights to reflect the influence of individual parameters and metrics, and (iv) consider dependencies, when computing risk scores.

\subsubsection{Quantification and Subjectivity} The proposed SAFER framework provided a clear and measurable way to formulate and express metrics. This quantification made it easier to judge the risk severity and to prioritize risks based on severity to make objective decisions. For instance, if a user has to choose between 2 software that can achieve the same objectives, one software with a risk score of 0.3 out of 1, whereas the other with a risk score of 0.55 out of 1, it will be easier for the user to choose among the two options. This also removed inconsistencies in risk computations performed by different parties. Each individual parameter (\emph{e.g.}, developer's reputation), major component (\emph{e.g.}, penalty), and risk segment (\emph{e.g.}, developer-based risk) has been represented in numeric value. All equations, \emph{i.e.}, equations~\ref{eq:riskcodedependencysimple}--\ref{eq:riskafterpenalty} yield quantitative values.

\subsubsection{Human Element} The devised framework integrated the risk associated with the human aspect of the software supply chain. It incorporated various parameters regarding individual actors, \emph{e.g.}, developer's expertise, publisher's experience, and user credibility. These parameters enabled a sensitive and comprehensive evaluation of the potential risks originating from the human elements throughout the software supply chain. The impact of human factors on risk has been formulated in equations~\ref{eq:totaldeveloperrisksimple},  \ref{eq:modifiedpubrisksimple}, and \ref{eq:modifieduserrecommrisk}.

\subsubsection{Data-driven Weights}
While developing the envisaged SAFER framework, data-driven weights have been employed instead of pre-defined static weights as evident from equations~\ref{eq:weightcodelength}, \ref{eq:weightforcodedependencyrisk}, \ref{eq:weightforcodespec}, etc.
The adaptability of dynamic weight allocation guarantees that the risk assessment remains relevant and reliable in the face of emerging threats and evolving operational environments by assessing a risk score based on the instantaneous values of the contributing parameters. In other words, the changes in the values of parameters used in the computation of weights associated with risk segments will also impact the influence of these risk segments in the final risk score. For instance, $R^{CD}$ is the risk based on code dependencies, and the weight associated with it is dependent on the code coverage which implies that if the code has been tested thoroughly (e.g., code coverage $C^C$ is 0.96), the weightage of the risk related to code dependencies $R^{CD}$ in the $R^{DEV}$ reduces (e.g., $w^{CD}$ is 0.04) and vice versa (e.g., $w^{CD}$ is 1 when code coverage $C^C$ is 0). Similarly, the higher the possibility of a vulnerability remaining unresolved, the greater the influence $w^{PB}$ of the publisher-based risk $R^{PB}$ in the $R^{F}$ and vice versa.

\subsubsection{Software Dependencies}
Software dependencies have been taken into account while formulating a risk element for developer-based risk, \emph{i.e.}, code dependencies-based risk in equation~\ref{eq:riskcodedependencysimple}. Moreover, the influence of dependencies on the risk based on code dependencies is adjustable according to the security requirements of the organization, as mentioned earlier in Section \ref{sec:dependecyimpact}.  

\subsection{Risk Assessments}

\begin{table}[t]
\begin{center}
\caption{Risk Computations \& Bands.}%
\label{tab:computations}
{\renewcommand{\arraystretch}{0.05}
\scriptsize	{
\begin{tabular}
{p{0.75cm}p{1.0cm}p{0.5cm}p{0.5cm}p{0.5cm}p{0.5cm}p{1.5cm}}
\toprule
\textbf{\centering{Sample}} & \textbf{\centering{$R^{\textup{Dev}}$}} & \textbf{\centering{$R^{\textup{Pb}}$}} & \textbf{\centering{$R^{\textup{ur}}$}}  & \textbf{\centering{$R^{\textup{F}}$}} & \textbf{\centering{$R^{\textup{FP}}$}} & 
\textbf{\centering{Band}}\\
\tabularnewline
\midrule
\centering{1} & \centering{48476.27} & \centering{0.37} & \centering{0.66} &  \centering{0.37} & \centering{0.37} & \centering{Moderate Risk}\\\tabularnewline

\centering{2} & \centering{225369.1} & \centering{0.04} & \centering{0.45}  & \centering{0.64} & \centering{0.65} & \centering{High Risk}\\\tabularnewline[-0.2cm]

\centering{~\vdots} & \centering{} & \centering{} & \centering{} & \centering{} & \centering{}  & \centering{~\vdots}\\\tabularnewline[0.3cm]

\centering{4500} & \centering{96020.01} & \centering{0.02} & \centering{0.78}  & \centering{0.95} & \centering{0.98} & \centering{Critical Risk}\\\tabularnewline

\centering{4501} & \centering{295440.1} & \centering{0.03} & \centering{0.53} & \centering{0.21} & \centering{0.21} & \centering{Low Risk}\\\tabularnewline[-0.2cm]

\centering{~\vdots} & \centering{} & \centering{} & \centering{} & \centering{} & \centering{} & \centering{~\vdots}\\\tabularnewline[0.3cm]

\centering{8999} & \centering{34403.97} & \centering{0.04} &  \centering{0.87} & \centering{0.79} & \centering{0.82} & \centering{Critical Risk}\\\tabularnewline

\centering{9000} & \centering{14369.99}  & \centering{0.02} &  \centering{0.78}  & \centering{0.40}  & \centering{0.40} & \centering{Moderate Risk}\\\tabularnewline[-0.2cm]
\bottomrule
\end{tabular}
}}
\end{center}
\end{table}

\begin{table}[t]
\begin{center}
\caption{User Experience Comparison.}%
\label{tab:userexperience}
{\renewcommand{\arraystretch}{0.05}
\scriptsize	{
\begin{tabular}
{p{2.5cm}p{0.5cm}p{0.5cm}p{0.5cm}p{0.5cm}p{0.5cm}p{0.5cm}}
\toprule
\textbf{\centering{Frameworks}} & \textbf{\centering\rotatebox[origin=c]{90}{Usability}} & \textbf{\centering\rotatebox[origin=c]{90}{Satisfaction}} & \textbf{\centering\rotatebox[origin=c]{90}{Efficiency}}  & \textbf{\centering\rotatebox[origin=c]{90}{Learnability}} & \textbf{\centering\rotatebox[origin=c]{90}{Accessibility}}\\
\tabularnewline
\midrule
\centering{OpenSSF Scorecard} & \centering{20\%} & \centering{100\%} & \centering{100\%} &  \centering{20\%} & \centering{20\%}\\\tabularnewline

\centering{OWASP Risk Calculator} & \centering{20\%} & \centering{20\%} & \centering{100\%} &  \centering{20\%} & \centering{100\%}\\\tabularnewline

\centering{Proposed SAFER} & \centering{100\%} & \centering{100\%} & \centering{100\%} &  \centering{80\%} & \centering{100\%}\\\tabularnewline[-0.2cm]

\bottomrule
\end{tabular}
}}
\end{center}
\end{table}

\begin{figure*}[!b]
\centering
\includegraphics[scale=0.70]
{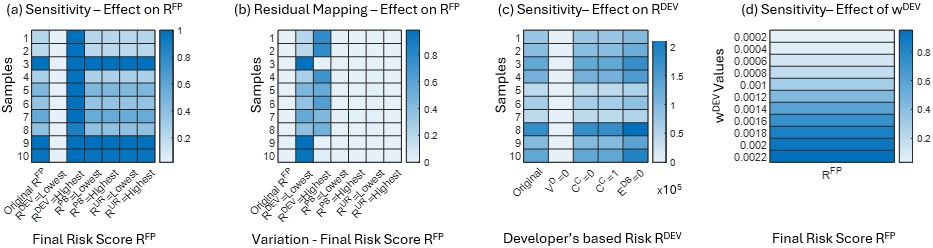}
\caption{Heat Maps - Sensitivity of the Proposed SAFER Framework (a) Effect of Individual Risk Segments on Final Risk Score, (b) Residual Mapping for Effect of Individual Risk Segments on Final Risk Score, (c) Effect of Individual Parameters on Developer based Risk, and (d) Effect of Changing Values of Developer based Risk Weightage on Final Risk Score.}
\label{fig:sensitivityheatmaps}
\end{figure*}

\begin{figure*}[t]
\centering
\includegraphics[scale=0.70]
{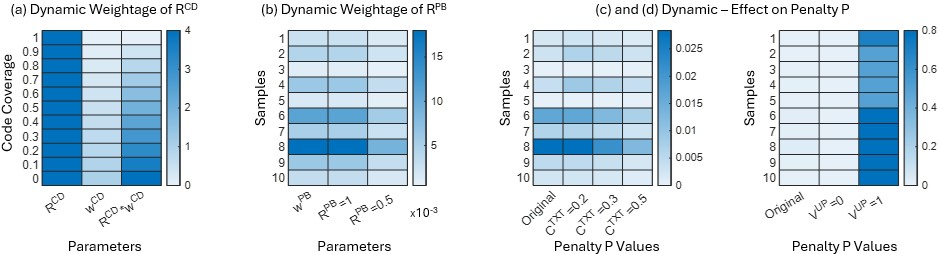}
\caption{Heat Maps - Dynamicity of the Proposed SAFER Framework (a) Effect of Changing Values of weight $w^{CD}$ associated with Code Dependencies based Risk on Contribution towards $R^{DEV}$, (b) Effect of Changing Values of Publisher based Risk $R^{PB}$ on Contribution towards $R^{F}$ where column 2 and column 3 represent the values of $w^{PB}*R^{PB}$, (c) Effect of Changing Context on Penalty, and (d) Effect of Changing Proportion of Unresolved Vulnerabilities on Penalty.}
\label{fig:dynamicityheatmaps}
\end{figure*}

The overarching observation in Table \ref{tab:computations} indicates a strong correlation between the overall risk ($R^F$) and the inherent risks associated with the involved developers ($R^{DEV}$) introduced during the development process and the risks indicated by users ($R^{UR}$). This linkage highlights the significance of delving into and proactively managing challenges that may arise at the development stage and considering the requirements of consumers. The importance of the reliance on developer-based risk on dependencies lies in the acknowledgment that the risk posed by dependencies is a crucial determinant in the overall risk profile associated with the development process. 



\subsubsection{Final Risk Score}
The risk associated with the developer for Sample 1 in Table \ref{tab:computations} is assessed as low, and the risk attributed to the publisher and the user for the same software reaches the moderate and the high level, respectively, primarily attributed to the publisher's lack of experience and a significantly lower number of users rating the software. Nevertheless, the impact of this high risk is mitigated in the final risk score $R^F$ due to a finely proportioned resolution of vulnerabilities.

The elevated developer risk in Sample 4501 arises from a heightened code specification-related risk, accompanied by a suboptimal proportion of vulnerability resolution. The software is designed for other applications (\emph{i.e.}, context) and, therefore, adheres to less stringent criteria for risk assessment.

\begin{table}[t]
\begin{center}
\caption{Comparison - OpenSSF Scorecard Risk Vs. Proposed SAFER Framework Risk.}%
\label{tab:comparisonossf}
{\renewcommand{\arraystretch}{0.05}
\scriptsize	{
\begin{tabular}{
p{1.2cm}p{1.5cm}p{1.2cm}p{1.5cm}p{1.3cm}}
\toprule
\textbf{\centering{Proposed Risk}} & \textbf{\centering{Proposed Band}} & \textbf{\centering{Scorecard Risk}}  & \textbf{\centering{Scorecard Band}} & \textbf{\centering{Manual Verification}}\\
\midrule
\centering{0.99} & \centering{Critical Risk} & \centering{1.8} & \centering{High Risk} & \centering{Critical
}\\\tabularnewline

\centering{0.27} & \centering{Moderate Risk} & \centering{7.3} & \centering{Low Risk} & \centering{Moderate
}\\\tabularnewline

\centering{0.22} & \centering{Low Risk} & \centering{4.1} & \centering{Medium Risk} & \centering{Low
}\\\tabularnewline

\centering{0.37} & \centering{Moderate Risk} & \centering{5.7} & \centering{Low Risk} &\centering{Moderate}\\\tabularnewline

\centering{0.2} & \centering{Low Risk} & \centering{5.3} & \centering{Low Risk} & \centering{Low}\\\tabularnewline

\centering{0.97} & \centering{Critical Risk} & \centering{2.3} & \centering{High Risk} &\centering{Critical}
\\\tabularnewline

\centering{0.39} & \centering{Moderate Risk} & \centering{5.7} & \centering{Low Risk} &\centering{Moderate}
\\\tabularnewline

\centering{0.46} & \centering{Moderate Risk} & \centering{2.0} & \centering{High Risk} & \centering{High
}\\\tabularnewline

\centering{0.56} & \centering{High Risk} & \centering{6.1} & \centering{Low Risk} &\centering{Moderate}\\\tabularnewline

\centering{1.0} & \centering{Critical Risk} & \centering{2.1} & \centering{High Risk} &\centering{Critical}\\\tabularnewline[-0.2cm]
\bottomrule
\end{tabular}
}}
\end{center}
\end{table}

\subsection{Evaluation \& Comparisons}
Table \ref{tab:userexperience} outlines the comparison of user experiences associated with the OpenSSF scorecard, OWASP risk calculator, and the proposed SAFER framework for risk assessment of private repositories. \textbf{Usability} refers to how easy and user-friendly the tool is to operate. Proposed SAFER: 100\% of the participants rated it as easy and user-friendly to operate. OWASP: non-expert users found selecting options via the dropdown menu to be quite easy, however, due to the lack of knowledge regarding the threat agents, vulnerability factors, and/or impact factors, struggled to operate the tool effectively. OpenSSF scorecard: none of the participants had GitHub advanced security license, therefore, they needed to operate it through the command line. Most participants, specially non-expert and non-technical participants found the installation process complicated and limited support for windows' platform posed an additional challenge. \textbf{Satisfaction} relates to how satisfied users are with the tool's performance and functionality. Proposed SAFER: 100\% of the participants rated it as satisfactory. OWASP: non-expert and non-technical participants were all able to compute risk scores using OWASP, however, they were not confident in the options they selected on individual dropdown menus. OpenSSF scorecard: As most of the participants couldn't get past the installation stage, they tried OpenSSF scorecard for public repositories and 100\% of the participants rated it as satisfactory. \textbf{Efficiency} indicates how quickly and effectively users can complete their tasks using the tool. Proposed SAFER: 100\% of the participants rated it as efficient. OWASP: 100\% of the participants rated it as efficient. OpenSSF scorecard: as mentioned before, majority of the participants were unable to use OpenSSF for private repositories, they rated the efficiency of OpenSSF scorecard by trying OpenSSF scorecard for public repositories and 100\% of the participants rated it as efficient. \textbf{Learnability} describes how easy it is for new users to learn how to operate the tool. Proposed SAFER: 80\% of the participants rated it as effortless. OWASP: non-expert and non-technical users found selecting options through the dropdown menu to be straightforward, however, found it challenging to understand the individual influencing parameters. OpenSSF scorecard: Majority of the participants were unable to even complete the installation stage, therefore, they couldn't start using OpenSSF scorecard for private repositories. \textbf{Accessibility} refers to how accessible the tool is to users with different abilities and needs. Proposed SAFER: 100\% of the participants rated it accessible. OWASP: 100\% of the participants rated it accessible. OpenSSF scorecard: all the participants were able to find all the files needed for the installation process, however, the OpenSSF scorecard for private repositories itself was inaccessible for them as they couldn't finish installation.

Figure \ref{fig:sensitivityheatmaps} demonstrates the sensitivity of the proposed SAFER framework for risk assessment. Figure \ref{fig:sensitivityheatmaps}(a) and (b) illustrate the effect of individual risk segments, i.e., $R^{DEV}$, $R^{PB}$, and $R^{UR}$ on the final risk score $R^{FP}$ via actual $R^{FP}$ value heat map and a residual heat map, respectively. It is evident that the risk segment associated with the development phase, i.e., $R^{DEV}$, has the most significant impact on determining the final risk value. This suggests that factors and uncertainties within the development stage play a crucial role in shaping the overall outcome, making it a critical area of focus for risk management and mitigation efforts. Similarly, Figure \ref{fig:sensitivityheatmaps}(c) presents the effect of the number of vulnerabilities in all the software developed by the same developer $V^{D}$, code coverage $C^{C}$, and experience of the developer in terms of programming language $E^{DB}$ on the developer's based risk segment $R^{DEV}$, whereas Figure \ref{fig:sensitivityheatmaps}(d) depicts the affect of the weight $w^{DEV}$ associated with the developer's based risk segment $R^{DEV}$ on the final risk score $R^{FP}$. It can be seen that the vulnerabilities in the developed software ($V^{D}$) has the most impact on the developer's based risk segment $R^{DEV}$ followed by the developer's expertise in the programming language ($E^{DB}$). It can also be observed that the increasing weight $w^{DEV}$ associated with the developer's based risk segment has an increasing impact on the final risk score $R^{FP}$.

Figure \ref{fig:dynamicityheatmaps} exhibits the dynamic aspect of the proposed SAFER framework for risk assessment. Figure \ref{fig:dynamicityheatmaps}(a) indicates the effect of varying values of weight $w^{CD}$ associated with code dependencies based risk on contribution towards $R^{DEV}$. It can be seen that the increasing code coverage has a decreasing impact on the weight associated with $R^{CD}$, consequently reducing the contribution of $R^{CD}$ towards $R^{DEV}$. Similarly, Figure \ref{fig:dynamicityheatmaps}(b) shows the impact of changing values of publisher-based risk $R^{PB}$ on contribution towards $R^{F}$. It is apparent that a reduced publisher-based risk while keeping the weight associated with it constant has a lower contribution towards the final score $R^{F}$ and vice versa. Figure \ref{fig:dynamicityheatmaps}(c) depicts the influence of distinct context on penalty, whereas Figure \ref{fig:dynamicityheatmaps}(d) presents the effect of changing the proportion of unresolved vulnerabilities on penalty. It can be noted that the penalty values applied on the security software (i.e., $C^{TXT}$ = 0.2) are much higher than the penalty on other software while keeping the proportion of unresolved vulnerabilities ($V^{UP}$) constant. In contrast, the increasing proportion of unresolved vulnerabilities ($V^{UP}$) while keeping the context constant has an increasing impact on the penalty.

\begin{table}[!b]
\begin{center}
\caption{Comparison - Proposed SAFER Framework Risk with/without Self-Generated Parameters.}%
\label{tab:comparisonwwo}
{\renewcommand{\arraystretch}{0.05}
\scriptsize	{
\begin{tabular}{p{1.7cm}p{1.7cm}p{1.7cm}p{1.7cm}}
\toprule
\textbf{\centering{Proposed Risk with $\mathcal{D}$}} & \textbf{\centering{Proposed Band with $\mathcal{D}$}} & \textbf{\centering{Proposed Risk with $\mathcal{D}_{x}$}}  & \textbf{\centering{Proposed Band with $\mathcal{D}_{x}$}}\\
\tabularnewline
\midrule

\centering{0.99} & \centering{Critical Risk} & \centering{0.98 } & \centering{Critical Risk}\\\tabularnewline

\centering{0.27} & \centering{Moderate Risk} & \centering{0.88} & \centering{Critical Risk}\\\tabularnewline

\centering{0.22} & \centering{Low Risk} & \centering{0.95} & \centering{Critical Risk}\\\tabularnewline

\centering{0.37} & \centering{Moderate Risk} & \centering{0.96} & \centering{Critical Risk}\\\tabularnewline

\centering{0.2} & \centering{Low Risk} & \centering{0.92} & \centering{Critical Risk}\\\tabularnewline

\centering{0.97} & \centering{Critical Risk} & \centering{0.95} & \centering{Critical Risk}\\\tabularnewline

\centering{0.39} & \centering{Moderate Risk} & \centering{0.94} & \centering{Critical Risk}\\\tabularnewline

\centering{0.46} & \centering{Moderate Risk} & \centering{0.97} & \centering{Critical Risk}\\\tabularnewline

\centering{0.56} & \centering{High Risk} & \centering{0.97} & \centering{Critical Risk}\\\tabularnewline

\centering{1.0} & \centering{Critical Risk} & \centering{0.98} & \centering{Critical Risk}\\\tabularnewline[-0.2cm]
\bottomrule
\end{tabular}}}
\end{center}
\end{table}

\begin{table*}[!b]
\centering
\caption{Risk Scores by OWASP Risk Calculator.}
\label{tab:riskowasp5}
{\renewcommand{\arraystretch}{1.2}
\footnotesize{
\begin{tabular}{ccccc}
\hline

\multicolumn{1}{c}{\textbf{Assessor 1}} & \multicolumn{1}{c}{\textbf{Assessor 2}} &  \multicolumn{1}{c}{\textbf{Assessor 3}}  &  \multicolumn{1}{c}{\textbf{Assessor 4}} & \multicolumn{1}{c}{\textbf{Assessor 5}}       \\ \hline
 
 \multicolumn{1}{c}{Low (LF:5.875, IF:2)}                & \multicolumn{1}{c}{Medium (LF:4.5, IF:3.75)} & \multicolumn{1}{c}{Medium (LF:6, IF:2.75)} & \multicolumn{1}{c}{Low (LF:5.875, IF:2)}  & \multicolumn{1}{c}{Low (LF4.375:, IF:2)}                                            \\ 
 
 \multicolumn{1}{c}{Low (LF:5.875, IF:2)}                & \multicolumn{1}{c}{Low  (LF:2, IF:3.75)} & \multicolumn{1}{c}{Medium (LF:6.125, IF:2.75)} & \multicolumn{1}{c}{Critical (LF:7.375, IF:7.5)} & \multicolumn{1}{c}{Medium (LF:5.125, IF:3.25)}                                         \\ 
 
 \multicolumn{1}{c}{Low (LF:5.625, IF:1.75)}                & \multicolumn{1}{c}{Medium (LF:4, IF:3.75)} & \multicolumn{1}{c}{Low (LF:5.25, IF:2.75)} & \multicolumn{1}{c}{Critical (LF:7.125, IF:7.5)}  & \multicolumn{1}{c}{Medium (LF:4.125, IF:3.5)}                                         \\ 

 \multicolumn{1}{c}{Medium (LF:5.625, IF:4.25)}                & \multicolumn{1}{c}{High (LF:6.125, IF:3.75)} & \multicolumn{1}{c}{Medium (LF:6.125, IF:2.75)} & \multicolumn{1}{c}{Critical (LF:7.5, IF:9)}  & \multicolumn{1}{c}{High (LF:3.25, IF:6)}                                             \\ 

 \multicolumn{1}{c}{Medium (LF:6, IF:2.25)}                & \multicolumn{1}{c}{Medium (LF:4.625, IF:3.75)} & \multicolumn{1}{c}{Low (LF:5.875, IF:2.75)} & \multicolumn{1}{c}{Medium (LF:6, IF:2.25)} & \multicolumn{1}{c}{Medium (LF:4.375, IF:5.75)}                                    \\ 

 \multicolumn{1}{c}{Low (LF:5.25, IF:2.75)}                & \multicolumn{1}{c}{High (LF:6.5, IF:3.75)} & \multicolumn{1}{c}{Low (LF:5.375, IF:2.75)} & \multicolumn{1}{c}{Low (LF:5.25, IF:2.75)}  & \multicolumn{1}{c}{Medium (LF:4.875, IF:5.5)}                                         \\ 

  \multicolumn{1}{c}{High (LF:6, IF:3.75)}                & \multicolumn{1}{c}{Low (LF:2.625, IF:3.75)} & \multicolumn{1}{c}{Medium (LF:7, IF:2.75)} & \multicolumn{1}{c}{Critical (LF:7.25, IF:9)} & \multicolumn{1}{c}{High (LF:3.875, IF:8)}                                            \\ 

  \multicolumn{1}{c}{Low (LF:5.25, IF:2.75)}                & \multicolumn{1}{c}{Medium (LF:3.125, IF:3.75)} & \multicolumn{1}{c}{Low (LF:5.5, IF:2.75)} & \multicolumn{1}{c}{Low (LF:5.25, IF:2.75)} & \multicolumn{1}{c}{Low (LF:5.75, IF:2)}                                          \\ 

 \multicolumn{1}{c}{Low (LF:5.25, IF:1.75)}                & \multicolumn{1}{c}{Medium (LF:5.375, IF:3.75)} & \multicolumn{1}{c}{Low (LF:5.625, IF:2.75)} & \multicolumn{1}{c}{Critical (LF:6.875, IF:7)} & \multicolumn{1}{c}{Low (LF:3.125, IF:2.5)}                                         \\ 

\multicolumn{1}{c}{Low (LF:5.875, IF:2)}                & \multicolumn{1}{c}{Low (LF:2.375, IF:3.75)} & \multicolumn{1}{c}{Medium (LF:6.125, IF:2.75)} & \multicolumn{1}{c}{High (LF:6, IF:5)}   & \multicolumn{1}{c}{Low (LF:5.25, IF:2)}                                       \\ \hline

\end{tabular}}}
\end{table*}

Table \ref{tab:comparisonossf} presents 10 random samples extracted from the results generated by utilizing the OpenSSF scorecard alongside the corresponding risk scores derived from the proposed risk assessment framework SAFER for the same repositories. While comparing the outcomes yielded by both methods, it was observed that the scores computed by the two are not directly correlated. Some scores computed by the OpenSSF scorecard suggest a higher risk band as compared to the risk scores computed by SAFER, whereas some recommend a lower risk band as opposed to the calculations made by using SAFER. This indicates that the parameters or risk factors not considered by the OpenSSF scorecard have an impact on the final risk scores. This implies that the attributes taken into account by the OpenSSF scorecard to assess the security state of the software are insufficient for accurately determining the risk level associated with the software. It necessitates additional parameters to be taken into account to properly identify the risk status and make it more stringent. Please note that the bands utilized by OpenSSF scorecard for final score are low ($5 \leq score \leq 10$), medium ($2.5 \leq score < 5$), high ($1.5 \leq score < 2.5$), critical ($score< 1.5$) represented by colored rings of yellow, orange, red, and maroon respectively on the scores, whereas the risk bands applied on the final risk score in the proposed methodology are low ($<0.25$), moderate ($0.25 \leq risk < 0.5$), high ($0.5 \leq risk < 0.75$), and critical ($0.75 \leq risk \leq 1$). Sample 3 has been scored as a medium risk by the OpenSSF scorecard (column 3) despite having no vulnerabilities, no dangerous workflow patterns, no defined security policy, etc. Similarly, Sample 9 has been scored as low risk by the OpenSSF scorecard even though vulnerabilities exist, the code review has less than 50\% changes approved, no fuzzing or static code analysis has been used, and there is no best practices badge associated. Proper formulation instead of pre-defined values for assigning scores and severity weight to individual parameters is equally as important as selecting the most relevant influencing parameters, the developer's and publisher's reputation and experience, along with user experience/satisfaction. 
Please note that the manual verification here includes all the parameters considered by the proposed risk framework, along with the the parameters employed by the OpenSSF scorecard. Sample 1 has the following: (1) no code coverage, (2) the developer has a reputation for developing software with a significantly high number of vulnerabilities, (3) the developer has little experience in developing software using this language, (4) no fuzzing, (5) no security policy, (6) no best practices, and (7) no information on dangerous workflows, (8) high number of detected vulnerabilities, (9) no branch protection, etc. Therefore, manual verification suggests that the risk level is critical. 

Similarly, Sample 8 has the following: (1) no code coverage, (2) the developer has a reputation for developing software with a moderate number of vulnerabilities, (3) the developer has little experience in developing software using this language, (4) the context of the software is other than security or automation, (5) no fuzzing, (6) no security policy, (7) no best practices, and (8) no information on dangerous workflows, etc. Accordingly, manual verification suggests that the risk level is high. Please note that the manual verification risk rating is not according to the number of quoted risk factors. It is according to the severity of the risk factors and the scoring against them. 

Table \ref{tab:comparisonwwo} presents the comparison of the risk scores calculated using the proposed SAFER framework by first employing the entire dataset $\mathcal{D}$ and then $\mathcal{D}_{x}$ is used which excludes the parameters (\emph{i.e.}, regarding the date of the software, experience in language, developer's and publisher's experience, number of downloads, update frequency, context, and penalty/reward) for which self-generated data points were utilized. While computing the risk scores for $\mathcal{D}_{x}$, equation~\ref{eq:modifieduserrecommrisk} was modified to remove the parameter for a number of downloads and normalize the value of user ratings to bring it into the range [0,1]. The values computed with all the parameters differ significantly from the ones calculated by employing only a subset of all the parameters. Moreover, the values computed by employing $\mathcal{D}_{x}$ (\emph{i.e.}, values in column 3) are exceptionally high, making all the software critically risky. Moreover, these values do not directly correlate to their counterparts, which demonstrates the eminence of the excluded parameters and reflects the importance of the security context of the software, the reputation of the developer and publisher, and the penalty to assess the trust-based risk level. The parameters impacting the values presented in column 3 include code dependencies, code coverage, user ratings, forks, vulnerabilities resolved, and total vulnerabilities.

Table \ref{tab:riskowasp5} delineates risk scores calculated by using the OWASP risk calculator by 5 assessors working in the computer systems and security domain and have been associated with research and engineering for 5 years or more. As mentioned before, results evaluated by the OWASP risk calculator are highly subjective (biased), and it is evident from the scores computed by different assessors in the same setting and context. Please note that $LF$ and $IF$ represent the likelihood factor and impact factor, respectively.

\section{Conclusion \& Future Work}
\label{sec:conclusion}
This paper has thoroughly explored the risk assessment associated with software security and shed light on the complex relationship of factors influencing the security and trust of software systems.  
By evaluating risk scores and categorizing them into bands, we have established a framework for comprehending and communicating the level of risk systematically without subjectivity. Exploring the human element via three key actors (\emph{i.e.}, developers, publishers, and users) and corresponding metrics in the software supply chain further enriched our understanding of the multifaceted nature of software risk.

In the future, the authors intend to carry out a qualitative study to rigorously assess quantitative risk assessment frameworks in the software supply chain. Another promising direction for ongoing investigation is developing a simulator for software supply chain simulation and data generation for risk assessment. Such a simulator will significantly enhance our ability to model and analyze diverse risk scenarios, enabling more comprehensive risk assessments. Through simulated data generation, we will be able to systematically test the resilience of software systems under various conditions, helping us refine risk quantification methodologies and fortify our defenses against potential vulnerabilities. Moreover, we intend to utilize techniques like reinforcement learning (RL) to evaluate whether RL can provide a more effective numeric score in addition to classification as well as facilitate the validation process. 

\section{Acknowledgments}
This research paper is conducted under the 6G Security Research and Development Project, as led by the Commonwealth Scientific and Industrial Research Organisation (CSIRO) through funding appropriated by the Australian Government’s Department of Home Affairs. This paper does not reflect any Australian Government policy position. For more information regarding this Project, please refer to \url{https://research.csiro.au/6gsecurity/}.

\balance

\section*{Appendix}
\textbf{Dataset}: The parameters included in the subset 
$\mathcal{D} - \mathcal{D}_{x}$ are the developer, publisher, year, language, update frequency, code length, and context. For the developer, 26 unique values represented by the 26 alphabets in the English language have been randomly generated. The same is the case with publishers, and there is no connection between the developers and the publishers. For the parameter year, the values from 2016 to 2023 were randomly selected. 3 different languages have been used here, including C, Python, and Java. Update Frequency is a random number between 0 and 1. Code length is a random number between 40 and 700. These numbers are selected by taking the number of code lines from two randomly selected repositories. The context is a random number selected from the 3 possible values of 0.2, 0.3, and 0.5. These values are selected to keep the sum of all equal to 1.

\begin{table*}[t]
\begin{center}
\caption{Sample Data for Example.}%
\label{tab:sampledataexample}
{\renewcommand{\arraystretch}{0.05}
\scriptsize{
\begin{tabular}{p{0.4cm}p{0.4cm}p{0.4cm}p{0.4cm}p{1.1cm}p{0.9cm}p{0.4cm}p{0.6cm}p{0.4cm}p{0.6cm}p{0.4cm}p{0.6cm}p{0.4cm}p{0.6cm}}
\toprule
\textbf{\centering\rotatebox[origin=c]{90}{Code Length}} & 
\textbf{\centering\rotatebox[origin=c]{90}{Developer}}  &
\textbf{\centering\rotatebox[origin=c]{90}{Publisher}}  &
\textbf{\centering\rotatebox[origin=c]{90}{Year}} & 
\textbf{\centering\rotatebox[origin=c]{90}{Language}} & \textbf{\centering\rotatebox[origin=c]{90}{Update Frequency}}&
\textbf{\centering\rotatebox[origin=c]{90}{Forks}}& 
\textbf{\centering\rotatebox[origin=c]{90}{Downloads}}& \textbf{\centering\rotatebox[origin=c]{90}{Unresolved Vulnerabilities}}& 
\textbf{\centering\rotatebox[origin=c]{90}{Known Vulnerabilities}}& 
\textbf{\centering\rotatebox[origin=c]{90}{Dependencies}}& 
\textbf{\centering\rotatebox[origin=c]{90}{Rating}}&
\textbf{\centering\rotatebox[origin=c]{90}{Code Coverage}}&
\textbf{\centering\rotatebox[origin=c]{90}{Context}}\\
\tabularnewline
\midrule

\centering{$l$} & \centering{$j$} & \centering{$x$} & \centering{$t_k$} & \centering{Language} & \centering{$F^{UP}$} & \centering{$f$} &  \centering{$N^{DS}$} & \centering{$V^U$} & \centering{$V^T$} & \centering{$D$} & \centering{$G^{US}$} & \centering{$C^C$}& \centering{$C^{TXT}$} \\\tabularnewline

\centering{304} & \centering{W} & \centering{Z} & \centering{2018} & \centering{Java} & \centering{0.08424} & \centering{2003} &  \centering{20455} & \centering{135} & \centering{7556} & \centering{28} & \centering{7153} & \centering{0.99}& \centering{0.2} \\\tabularnewline[-0.2cm]
\bottomrule
\end{tabular}
}}
\end{center}
\end{table*}

\addtocounter{table}{-1}

\begin{table*}[t]
\begin{center}
\caption{Sample Data for Example (Continuation).}%
\label{tab:sampledataexample}
{\renewcommand{\arraystretch}{0.05}
\scriptsize{
\begin{tabular}{p{0.8cm}p{0.6cm}p{0.6cm}p{0.6cm}p{0.6cm}p{0.6cm}p{0.6cm}p{1.1cm}p{0.7cm}}
\toprule
\textbf{\centering\rotatebox[origin=c]{90}{$V^T$ in all the software developed by $j$}}&
\textbf{\centering\rotatebox[origin=c]{90}{Number of all software developed by $j$}}&
\textbf{\centering\rotatebox[origin=c]{90}{Experience of $j$ in terms of the number of years in the same language}}&
\textbf{\centering\rotatebox[origin=c]{90}{Total number of years $j$ has been developing software}}&
\textbf{\centering\rotatebox[origin=c]{90}{Number of previously developed software by
$j$ in the same language}}&
\textbf{\centering\rotatebox[origin=c]{90}{Experience of $x$ in terms of the number of software published}}&
\textbf{\centering\rotatebox[origin=c]{90}{Experience of $x$ in terms of years}}&
\textbf{\centering\rotatebox[origin=c]{90}{Proportion of Unresolved Vulnerabilities}}&
\textbf{\centering\rotatebox[origin=c]{90}{Resolved Vulnerabilities}}\\
\tabularnewline
\midrule

 \centering{$V^D$} & \centering{$|S^D|$} & \centering{$E^{DS}$} & \centering{$E^{DA}$} & \centering{$|S^L|$} & \centering{$E^{PX}$} & \centering{$E^{PY}$} & \centering{$V^{UP}$} & \centering{$V^R$} \\\tabularnewline

 \centering{96912} & \centering{129} & \centering{2} & \centering{2} & \centering{33} & \centering{123} & \centering{2} & \centering{0.017867} & \centering{7421} \\\tabularnewline[-0.2cm]
\bottomrule
\end{tabular}
}}
\end{center}
\end{table*}

\textbf{Repositories}: The repositories used in Tables \ref{tab:comparisonossf} and \ref{tab:riskowasp5} include (i) senchalabs/connect, (ii) pouchdb/pouchdb, (iii) webcomponents/webcomponentsjs, (iv) tj/commander.js, (v) thoughtbot/neat, (vi) openpgpjs/openpgpjs, (vii) webpack-contrib/sass-loader, (viii) webdriverio/webdriverio, (ix) jneen/parsimmon, and (x) rackt/redux-simple-router.

\textbf{Example}: The context value of 0.2 represents a security software. The values such as $V^D$ have been calculated considering all 9000+ samples.
\begin{align*}
    & V^{UP} = \frac{V^U}{V^T} \\
    & V^R = V^T - V^U
\end{align*}

\noindent
Equation 1:\\
$R^{\textup{CD}}_{S_i,t_k}= 28$\\

\noindent
Equation 3:\\
$w^{\textup{LC}}_{t_k} = \frac{96912}{129} = 751.2558$\\

\noindent
Equation 2:\\
$R^{\textup{CS}}_{S_i,t_k}= 751.2558*304 = 228381.8$\\

\noindent
Equation 4:\\
$R^{\textup{PL}}_{S_i,t_k}= 1- \frac{2}{2} = 0$\\

\noindent
Equation 6:\\
$w^{\textup{CD}}_{S_i,t_k} = 1-0.99 = 0.01$\\

\noindent
Equation 8:\\
$E^{\textup{DB}}_{S_i,t_k}=\frac{33}{129}= 0.255814$\\

\noindent
Equation 7:\\
$w^{\textup{CS}}_{S_i,t_k}= e^{-0.255814} = 0.77428$\\

\noindent
Equation 9:\\
$w^{\textup{PL}}_{S_i,t_k} = |1 - (0.01+0.77428)|= 0.2157$\\

\noindent
Equation 5:
\begin{flalign*}
R^{\textup{DEV}}_{S_i,t_k} &=0.01*28 + 0.77428*228381.8 + 0.2157*0 &\\
&= 176,831.46 &
\end{flalign*}

\noindent
Equation 10:\\
$R^{\textup{PB}}_{S_i,t_k}= \frac{2}{123} * \frac{1}{0.08424}=0.1930$\\

\noindent
Equation 11:\\
$R^{\textup{UR}}_{S_i,t_k} = 1- \frac{7153}{20455} = 0.6503$\\

\noindent
Equation 13:\\
$C^{\textup{TXT}} = 0.2$\\

\noindent
Equation 12:\\
$P_{S_i,t_k} = 1-({0.2})^{0.017867}= 0.0283$\\

\noindent
Equation 17:\\
$w^{\textup{DEV}}_{S_i,t_k} = \frac{1}{2003} = 4.99 * 10^{-4}$\\

\noindent
Equation 18:\\
$w^{\textup{PB}}_{S_i,t_k} = \frac{135 + 1}{7556 + 2} = 0.0179$\\

\noindent
Equation 16:\\
$w^{\textup{UR}}_{S_i,t_k} = 1 - (4.99 * 10^{-4} + 0.0179) = 0.9815$\\

\noindent
Equation 15:
\begin{align*}
R^{\textup{F}}_{S_i,t_k} &= \frac{1}{1+e^{4-0.04(0.00049 \cdot 176,831.46 + 0.0179 \cdot 0.1930 + 0.9815 \cdot 0.6503)}} \\
&= 0.3755
\end{align*}

\noindent
Equation 19:\\
$R^{\textup{FP}}_{S_i,t_k} = R^{\textup{F}}_{S_i,t_k} = 0.3755$\\

\textbf{Tests - Errors and Missing Information:}
Following tests related to errors and missing information against parameters have been performed, and assumptions are embedded in the functionality of the framework for added convenience. 

\begin{enumerate}
    \item Do not put any value for dependencies
    . Did you get $R^{\textup{FP}}$ without error
    ?
    \item Put value for code coverage greater than 1
    . Did you get an error?
    \item Put value for context other than 0.2, 0.3, or 0.5
    . Did you get an error?
    \item Leave any 5 values/cells empty. Did you get $R^{\textup{FP}}$ without error
    ?
    \item Put data for 3 developers in the form 3,5,10
    . Did you get $R^{\textup{FP}}$ without error
    ?
    \item Put a decimal value for the number of downloads
    . Did you get an error message?
\end{enumerate}

\end{document}